\documentclass[prx,aps,graphics,a4paper,10pt,twocolumn,nofootinbib,longbibliography]{revtex4-1}

\usepackage[margin=25mm]{geometry}
\usepackage{amsmath}
\usepackage{amssymb}
\usepackage{braket}
\usepackage{hyperref}
\usepackage{pgfplots}
\usepackage{bbm}
\usepackage{siunitx}
\usepackage{array}

\newcommand{\new}[1]{{#1}}
\newcommand{\ra}[1]{\renewcommand{\arraystretch}{#1}}

\AtBeginDocument{%
    \newwrite\bibnotes
    \def\bibnotesext{Notes.bib}
    \immediate\openout\bibnotes=\jobname\bibnotesext
    \immediate\write\bibnotes{@CONTROL{REVTEX41Control}}
    \immediate\write\bibnotes{@CONTROL{%
    apsrev41Control,author="08",editor="1",pages="1",title="0",year="1"}}
     \if@filesw
     \immediate\write\@auxout{\string\citation{apsrev41Control}}%
    \fi
}%

\newcommand{\be}{\begin{equation}} 
\newcommand{\ee}{\end{equation}}
\newcommand{\ba}{\begin{array}}
\newcommand{\ea}{\end{array}}
\newcommand{\bqa}{\begin{eqnarray}}
\newcommand{\eqa}{\end{eqnarray}}

\newcommand{\data}{\mathcal{D}}

\newcommand{\ns}{M} 
\newcommand{\nr}{N} 
\newcommand{\nq}{N_{r}} 
\newcommand{\nsites}{L} 
\newcommand{\nbosons}{N_b} 
\newcommand{\np}{K} 
\newcommand{\nparams}{\new{S}} 
\newcommand{\gp}{Gaussian process } 
\newcommand{\bo}{Bayesian optimization } 
\newcommand{\fglobal}{f}

\newcommand{\matel}[3]{\langle #1|#2|#3\rangle}
\newcommand{\um}{\mathbbm{1}}

\begin{document}

\title{Optimal quantum control with poor statistics}

\author{Fr\'ed\'eric Sauvage}
\author{Florian Mintert}

\affiliation{Physics Department,	Blackett Laboratory, Imperial College London, Prince Consort Road, SW7 2BW, United Kingdom}
\date{\today}

\begin{abstract}
Control of quantum systems is a central element of high-precision experiments and the development of quantum technological applications.
Control pulses that are typically temporally or spatially modulated are often designed based on theoretical simulations.
As we gain control over larger and more complex quantum systems, however, we reach the limitations of our capabilities of theoretical modeling and simulations,
and learning how to control a quantum system based exclusively on experimental data can help us to exceed those limitations.\\
Due to the intrinsic probabilistic nature of quantum mechanics, it is fundamentally necessary to repeat measurements on individual quantum systems many times in order to estimate the expectation value of an observable with good accuracy.
Control algorithms requiring accurate data can thus imply an experimental effort that negates the benefits of avoiding theoretical modeling.
We present a control algorithm based on Bayesian optimization that finds optimal control solutions in the presence of large measurement shot noise and even in the limit of single-shot measurements.
\new{With several numerical and experimental examples we demonstrate that this method is capable of finding excellent control solutions with minimal experimental effort.}
\end{abstract}

\maketitle
\section{Introduction}
\label{sec:introduction}
Quantum control allows us to perform highly accurate experiments and to turn synthetic quantum systems into devices ranging from the most precise available clocks~\cite{Katori2011} to qubits registers~\cite{Monroe1164,Barends2014,Gambetta2017} on which several elementary quantum gates can be executed.
Optimization of the typically time-dependent control fields is traditionally performed based on a theoretical model.
Given the increasingly complex quantum systems that we are gaining control over, it often becomes infeasible to find a sufficiently accurate model, let alone to simulate its dynamics.
The design of control fields based on experimental observations, but without resort to theoretical modeling~\cite{PhysRevLett.68.1500} has thus become more and more important.

While it is generally important that control algorithms converge quickly to a good solution, optimal control based on experimental data also requires that such solutions can be found using a small amount of data and data with limited accuracy.
This is particularly problematic with modern experiments on quantum systems since the vast majority are performed on individual systems, rather than ensembles as was the case in the early days of quantum mechanics.
Since obtaining expectation values of an elementary observable with good accuracy requires a large number of repetitions of the same experiment,
there is a trade-off in the acquisition of data.
More repetitions yield data with higher accuracy, which is clearly helpful in the search for a good control pulse.
The repetitions, however, also imply additional effort, which may even make a search prohibitively expensive.

Our present goal is to devise a control algorithm that does not require accurate estimation of expectation values, and that can find good solutions of control problems even in the extreme case of poor statistics due to intrinsic shot noise.

Among the most popular methods to perform optimizations directly based on experimental outcomes, exact or approximate gradient-based routines \cite{PhysRevA.91.052306,li2017hybrid,npj.45,Dive2018insituupgradeof,feng2018gradient} and Nelder-Mead~\cite{rosi2013fast, van2016optimal, kelly2014optimal,poggiali2018optimal,egger2014adaptive} have the advantage of simplicity and fast convergence when the underlying optimization landscape is well-behaved.
Their performance can however be limited in the presence of many local minima \cite{PhysRevA.90.032310,bukov2018reinforcement} or plateaus \cite{mcclean2018barren}.
Evolutionary algorithms \cite{assion1998control,PhysRevA.90.032310,PALITTAPONGARNPIM2017116}, and the recently introduced reinforcement learning techniques \cite{Melnikov1221,bukov2018reinforcement,PhysRevX.8.031084,august2018taking,niu2019universal}
may overcome this problem at the expense of a large number of iterations that require a large set of data.

Bayesian optimization \cite{brochu2010tutorial,snoek2012practical,shahriari2015taking,frazier2018tutorial}, on the other hand, has been shown to offer a good compromise between the ability to find high-quality solutions quickly, based on limited, potentially noisy data and reliability in distinguishing local from global maxima~\cite{wigley2016fast,2017arXiv171205771O,Henson2018,zhu2018training,nakamura2019non,2020arXiv200103520M}.
Since at its heart lies probabilistic modeling we deem it an excellent starting point for our present purpose.

In this paper we thus develop a framework based on \bo which takes into account the specific properties of measurement noise,
and we will show its ability to convergence substantially faster than other available methods.
In section Sec.~\ref{sec:algorithm} we will review pertinent elements of \bo and develop the detailed modeling of measurement noise that will enable optimal control based on extremely noisy data.
\new{Specifically, Sec.~\ref{sec:surmodel}, \ref{sec:inpractice} and \ref{sec:decisionrules} are devoted to an introduction to Bayesian optimization,
while the explicit modeling of measurement noise is discussed in Sec.~\ref{sec:measnoise}, followed by a discussion of the formulation of control targets in Sec.~\ref{sec:targets}.
The explicit evaluation of predictions of Bayesian inference in the present framework is detailed in \ref{sec:integrals}.\\}
The discussion in Sec.~\ref{sec:atwork} of the performance of this framework with a pedagogical example of a single qubit,
the preparation of a GHZ state,
the realization of ordered states of ultra-cold gases
\new{and the state preparation in a NISQ device}
might be appreciated also by readers who decided to skip details of the discussion in Sec.~\ref{sec:algorithm}.

\section{Optimal control algorithm}
\label{sec:algorithm}

Optimal control~\cite{PhysRevA.37.4950,Werschnik_2007,Glaser2015} is applicable in any situation in which a system can be manipulated with external control fields, as realized for example by a laser- or micro-wave pulse.
The central task is the identification of the control pulse that achieves a desired goal as accurately as possible.
A figure of merit that characterizes how well such a goal has been achieved is often called target functional.
Typical examples for such figures of merit include state fidelities~\cite{PhysRevA.84.022326}, gate fidelities~\cite{dolde2014high,kelly2014optimal,banchi2016quantum}, the expectation values of an entanglement witness, or also more non-linear quantities like entanglement measures~\cite{Melnikov1221,PhysRevLett.105.020501,1367-2630-17-9-093014,PhysRevA.84.022326} and the Fisher information~\cite{PhysRevA.96.032310,poggiali2018optimal}.

In order to turn an optimization task into a practical problem, it is typically required to characterize control pulses in terms of a finite number,  \new{{\it i.e.} a vector $\theta$ of tuneable control parameters.}
Those are often given in terms of piecewise constant amplitudes of external electric or magnetic fields~\cite{Dive2018insituupgradeof,PhysRevA.91.052306,feng2018gradient,li2017hybrid,npj.45,Khaneja2005296},
but also temporal shapes of laser- or microwave pulses with parametrization in terms of Fourier series~\cite{PhysRevLett.115.190801,PhysRevLett.120.150401,rosi2013fast,van2016optimal,PhysRevA.84.022326,2019arXiv190505721O},
or rotation angles of quantum gates~\cite{peruzzo2014variational,farhi2014quantum} are objects of optimization.

In control based on theoretical modeling, the control parameters are tuneable parameters in the Hamiltonian or quantum gate,
but in control based on experimental data the control parameters would typically be inputs to an experimental device.
Crucially, it is not necessary to know how these inputs relate to the Hamiltonian describing the system, and even systems whose microscopic workings are not understood are accessible to this type of control.

The dependence of the figure of merit $F$ on the control parameters $\theta$ defines the control landscape,
and the task at hand is identifying the value of $\theta$ for which $F(\theta)$ adopts its maximum.
In most practical problems, however, one is often satisfied with the identification of a value of $\theta$ that results in a sufficiently large value of $F(\theta)$.

It is hardly ever possible to find the dependence of the figure of merit $F$ on the control parameters $\theta$ analytically.
In addition to the necessity of an underlying theoretical model, this would require constructing the solution of the Schr\"odinger equation as an analytic function of the control parameters, which is only feasible in exceptionally elementary situations.
Typically, one thus probes
 the value of $F(\theta)$ only for a discrete set of data points,
and it is essential to use this limited information wisely in order to gauge where the true maximum or at least a particularly large value of the control landscape can be found.

\new{Denoting the vectors of control parameters for which the control landscape has been probed by} $\theta_j$ ($j=1,\hdots,M$)
and denoting the corresponding observations by $\data=[F(\theta_1),\hdots,F(\theta_M)]$
this task can be formalized as:
\begin{quote}
given the observations $\data$,
what control pulse $\theta_{M+1}$ is expected to yield a particularly large value of $F$?
\end{quote}

\begin{figure}
	\includegraphics[width=0.48\textwidth]{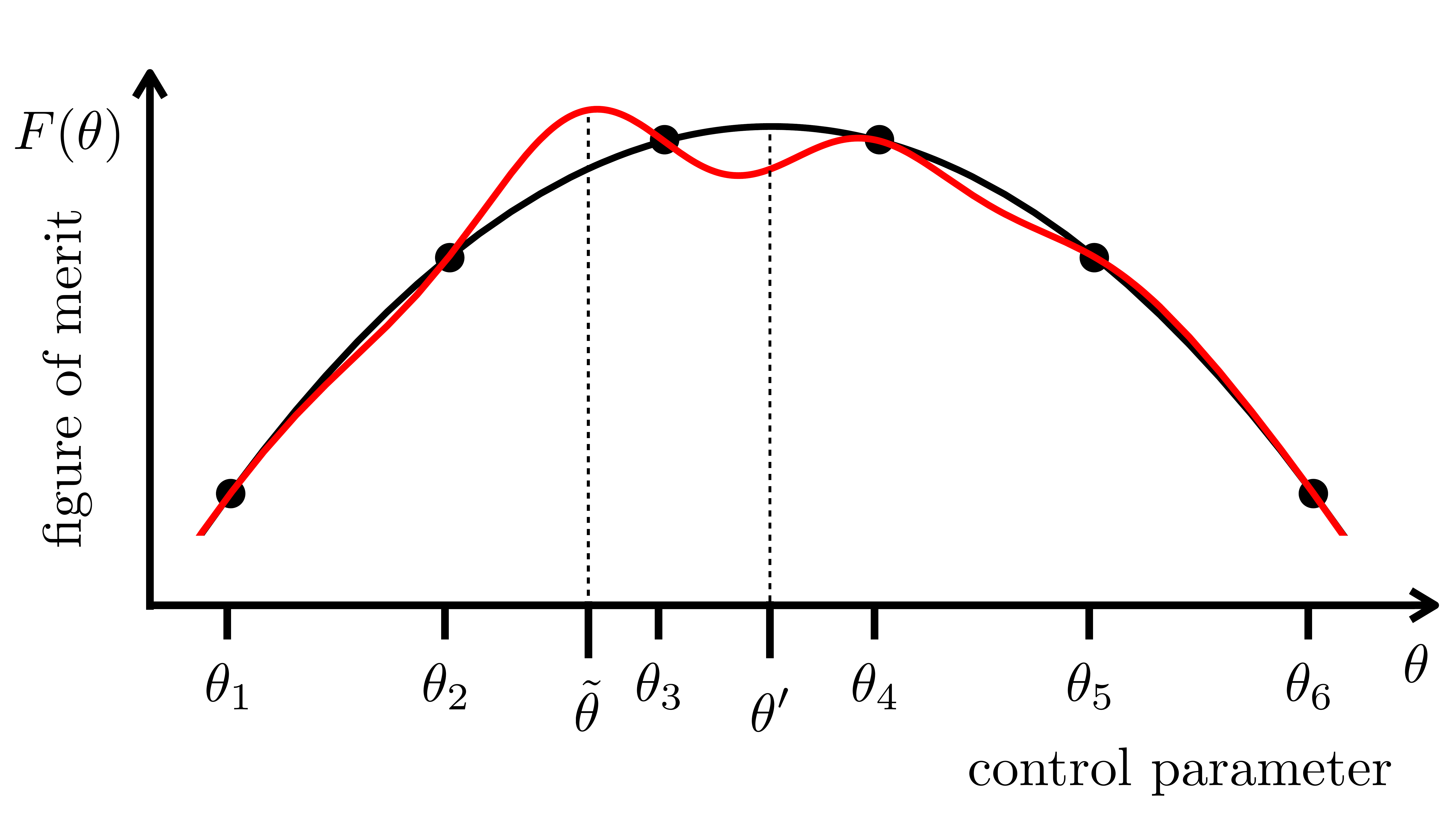}
	\caption{
	Schematic example of the estimation of a control landscape based on a finite number of data points.
	In this example, the six given data points are consistent with a parabola shown in black, but they are also consistent with many other landscapes, such as the one shown in red.
	Assuming the black landscape to be the correct one, one would predict the control parameter $\theta^\prime$ to yield the maximum of the landscape, but one would predict the control parameter $\tilde\theta$, assuming the red landscape to be the correct one.
	Even though the red landscape might not seem likely to be the correct one given the six data points, it can be worth considering since it promises a larger value of $F$ than the black landscape.
	}
	\label{fig:schematic}
\end{figure}

In practice, this question can be interpreted in different ways:
one may strive for a control pulse for which one is highly confident that it will yield a good result;
or, one may strive for a more optimistic approach, testing a control pulse that holds the potential to yield an exceptional result even if one is not confident that this will actually be the case, as sketched in Fig.~\ref{fig:schematic}.

Bayesian optimization provides the solid statistical footing for taking this decision.
It allows us to consider the full spectrum of conceivable control landscapes that are consistent with the given experimental observations.
In the spirit of Bayesian analysis, one can assign a probability to every landscape that characterizes one's belief for this landscape to coincide with the actual control landscape.
Since this provides a full probability distribution for conceivable control landscapes,
it not only allows us to construct an expected control landscape, but also to estimate when deviations from the expected behavior are sufficiently likely to justify exploring control parameters with high risk, but also high gain.

The ability to make predictions on the control landscape is then the basis for an iterative procedure.
Based on the available information one selects the most promising control pulse.
Its performance then needs to be assessed experimentally.
With the additional knowledge on the performance of this new pulse, one is able to make a more educated prediction for a subsequent pulse, and
the repetition of these steps is likely to result in the identification of a pulse that achieves the desired goal.

\subsection{Surrogate model} 
\label{sec:surmodel}

Since no finite number of data points will be enough to reconstruct the complete control landscape $F(\theta)$, Bayesian optimization aims at approximating this landscape in terms a set of conceivable control landscapes that are called proxy or surrogate models.
A simple example of such a surrogate model is given in Fig.~\ref{fig:schematic},
where the black parabola represents that actual -- but unknown -- control landscape,
and the red curve represents one possible surrogate model.
In practice one typically considers a continuous set of surrogate models that are consistent with the data obtained from the experiment.

The central quantity required to make statistically sounds predictions on the actual control landscape,
is the probability distribution for the actual control landscape to be given by a certain surrogate model $f$.
Since this shall be done based on prior experimental observations, this is given by the conditional probability distribution
\be\label{eq:pred}
P(f|\data)
\ee
for $f$, given the set of observations $\data$ made on the experiment.

The construction of this predictive probability density follows Bayes' basic rule
\be
P(f|\data)=\frac{P(f)P(\data|f)}{P(\data)}\ ,
\label{eq:post}
\ee
in terms of the prior probability distribution $P(f)$ of surrogate models,
the conditional probability distribution $P(\data|f)$ to obtain the acquired data for a given control landscape $f$, and
the probability $P(\data)$ to acquire the data $\data$ in a given sequence of experiments;
the latter can be understood as a normalization constant and can be obtained from the condition $\int df\ P(f|\data)=1$.

\subsubsection{The prior distribution}
\label{sec:prior}

The prior distribution $P(f)$ describes our limited knowledge on the control landscape before making any observations.
In the absence of detailed knowledge about a given system, it is chosen based on general, physically reasonable assumptions:
in practice, one would tend to find a landscape that varies smoothly with the control parameters more likely than a landscape with excessive oscillations or even discontinuities.

In the asymptotic limit, {\it i.e.} after infinitely many observations, Bayesian inference will identify the correct landscape essentially independently of the specific choice of prior distribution.
There is thus nothing like a unique, correct prior distribution, but we are free in its choice.
For \bo to find a good solution quickly, however, it is important to use a reasonable prior distribution that is sufficiently broad to contain the actual control landscape, but that does not assign high probability to landscapes with unnatural features such as excessively wild oscillations or close-to-discontinuous dependencies.
We defer the details of the practical definition of $P(f)$ to Sec.~\ref{sec:gaussian},
and take in the following for granted that we can actually specify such a distribution.

\subsubsection{Expected observations}

The most relevant term in Eq.~\eqref{eq:post} for our present purposes is the conditional probability density $P(\data|f)$.
Even though, one hardly ever knows the true control landscape, one can characterize accurately what observations one would expect with a given control landscape.

In a perfectly idealized, noiseless situation, one would expect to observe the true value $f(\theta)$ of the underlying landscape $f$ in an experiment performed with the control pulse $\theta$.
In this case, the conditional probability density for the figure of merit to adopt the value of $x$ would thus be given by $P(x|f)=\delta(x-f(\theta))$.

In a more realistic situation, any type of imperfection, such as finite resolution in a measurement or experimental noise can be incorporated in the term $P(\data|f)$, and instead of the above infinitely sharply peaked distribution, one obtains a probability distribution with a finite width.
This can be done for essentially any type of imperfection, or combinations therefore,
but we will develop it in more detail for the case of measurement noise in the following.

\subsection{Measurement noise} 
\label{sec:measnoise}

In the presence of large shot noise, it is helpful to express a figure of merit in terms of probabilities of measurement outcomes,
rather than expectation values of observables.
\new{In most practical situations, the figure of merit will not just be the probability to obtain a given outcome for one specific measurement,
but it will be a function of several physical observables, or, equivalently, a function of probabilities of outcomes of several measurements.
Since all these cases require the proper description of individual measurements,
we will elaborate on the modeling of a single observable
first, and defer the discussion of control targets comprised of several observables to Sec.~\ref{sec:targets}.}

\new{Since
the probability to obtain a given measurement outcome can only adopt values in the interval between $0$ and $1$,
one should impose the same condition on the surrogate model for this probability}.
For any surrogate model $f$ satisfying this condition, the conditional probability to obtain a detector click given the control landscape $f$ is given by
\be
P(1|f(\theta))=f(\theta)\ ,
\ee
whereas the probability to obtain no detector click is given by
\be
P(0|f(\theta))=1-f(\theta)\ .
\ee

This is readily generalized to the case of $\nr$ repetitions of the \new{same} measurement,
in which the probability to obtain $n$ detector clicks is given by the binomial distribution
\be
P(n|f(\theta))=\left(\nr \atop n\right)f(\theta)^{n}(1-f(\theta))^{\nr-n}\ .
\label{eq:binomial}
\ee

The estimation of a control landscape is not based on observations with one single control pulse $\theta$, but rather based on observations with several
\new{different control pulses $\theta_j$ utilized in different repetitions of the experiment.
Since for any given surrogate model $f$ the measurement outcomes in experiments performed with different control pulses are statistically independent,
the probability $P(\{n_j\}|f)$ to observe a sequence of $n_1$ to $n_{\ns}$ detectors clicks in a sequence of experiments performed with the  control pulses $\theta_1$ to $\theta_{\ns}$,
is given by the product}
\be
P(\{n_j\}|f)=\prod_{j=1}^{\ns} P(n_j|f(\theta_j))
\label{eq:PDf}
\ee
\new{of the probabilities $P(n_j|f(\theta_j))$ to obtain $n_j$ detector clicks in the experiments performed with the control pulse $\theta_j$.}

This long product of binomial distributions is the explicit form of the general object $P(\data|f)$ entering the Bayesian inference above in Eq. ~\eqref{eq:post}.
With this modeling in hand one can explicitly construct the predictive probability density $P(f(\theta)|\data)$,
based on any level of measurement noise, including the extreme case of no repetitions, {\it i.e.} $\nr=1$.

\subsection{Predictions}
\label{sec:inpractice}

With everything laid out so far, we are now ready to address the central step of predicting properties of the control landscape.
As argued above, this will always be based on a discrete set of data points obtained from a discrete set of control pulses $\theta_j$ (with $j=1,\hdots,\ns$).
It is essential that properties of the control landscape can be predicted for any conceivable control pulse $\theta$.

Since the goal is to do so for points corresponding to pulses that have not been experimentally explored yet,
it is helpful to first consider the conditional probability density
\be
P(f(\theta)|\vec f)
\label{eq:int1}
\ee
for the control landscape to contain the point $f(\theta)$,
given that it contains the points $\vec f=[f(\theta_1),\hdots,f(\theta_\ns)]$.
This is a quantity that can be characterized without any experimental observation, as it is given by
\be
P(f(\theta)|\vec f)=
\frac{P(f(\theta_1)\hdots,f(\theta_{\ns}),f(\theta))}{P(f(\theta_1)\hdots,f(\theta_{\ns}))}
\ee
in terms of the prior distribution $P(f)$ described above in Sec.~\ref{sec:prior}.
Once the prior is defined explicitly, the conditional density $P(f(\theta)|\vec f)$ is thus too.

Experimental observations are necessary in order to assess the probability for the actual control landscape to contain the points $\vec f$.
Due to the noise in the data acquisition, this can indeed only be done in terms of a probability distribution, and it is given exactly by $P(\vec f|\data)$.
This posterior distribution can be obtained according to Bayes' rule as
\be
P(\vec f|\data)=\frac{P(\vec f)P(\data|\vec f)}{P(\data)}.
\label{eq:int2}
\ee

With the probabilities $P(f(\theta)|\vec f)$ and $P(\vec f|\data)$, 
one can finally recast the desired predictive probability as
\be
P(f(\theta)|\data)=\int d^\ns\hspace{-1mm} f\ P(f(\theta)|\vec f)P(\vec f|\data)\ ,
\label{eq:predprob}
\ee
where $d^\ns\hspace{-1mm} f$ denotes the $\ns$-dimensional integral over the components of $\vec f$.

This is exactly the quantity that enables predictions for any point on the landscape before any experimental data resulting from the pulse corresponding to this point is available.
As anticipated above, it is not simply the expected value of the control landscape, but it is the full probability distribution, that allows one to estimate whether there is a sufficiently high chance to find a particularly high value of the control landscape for any new control pulse.

\subsection{Target functionals}
\label{sec:targets}

So far, we have restricted the discussion to the case in which the figure of merit is exactly the probability for a detector click.
In practice, however, many control targets $F$ depend on the outcomes of several measurements.
They can thus be expressed as function
\be
F = {\cal Q}(p_1, \hdots, p_K)
\label{eq:Fp}
\ee
of the probabilities $p_k$ for the outcomes of those measurements.

In this case, one can employ an independent surrogate model $f_k$ for each of the $\np$ required probabilities,
such that each $f_k$ can be used to make predictions for the probability $p_k$.
The surrogate model $\fglobal$ of the control target $F$, is then expressed as
\be
\fglobal  = {\cal Q}(f_1, \hdots, f_K),
\label{eq:Ff}
\ee
with exactly the same functional dependence as in Eq.~\eqref{eq:Fp}.
With this, one can construct the overall predictive distribution $P(f(\theta)|\data)$ for the actual control target
based on each individual predictive probability distributions $P(f_k(\theta)|\data)$ for the individual probabilities.

\subsection{Decision rules}
\label{sec:decisionrules}

With the probabilistic modeling for the control landscape defined,
the remaining question is how to use the predictive probability distribution $P(f(\theta)|\data)$ (Eq.~\eqref{eq:predprob}) in order to decide which control pulse to use in the next step of the iterative optimization.

At any level of given information obtained from the experiment can the expected control landscape be expressed as
\be
\langle \fglobal(\theta) \rangle=\int d\fglobal \fglobal(\theta) P(\fglobal(\theta)|\data)\ .
\label{eq:expected}
\ee
Bearing in mind that the symbol $P(f(\theta)|\data)$ is just a compact notation for
$P(f(\theta)=p|\data)$.
Since the probability $p$ lies in the interval $[0,1]$, Eq.~\eqref{eq:expected} can also be expressed as
\be
\langle \fglobal(\theta) \rangle=\int_0^1 dp\ p\ P(f(\theta)=p|\data)\ ,
\ee
which is a regular expectation value with respect to the probability distribution $P(f(\theta)|\data)$.

Analogously, also the variance
\be
\sigma^2(\theta)=\langle\left(\fglobal(\theta)-\langle \fglobal(\theta) \rangle\right)^2 \rangle
\ee
around the expected landscape, or any other function $A$ of $\fglobal(\theta)$ can readily be obtained in terms of a regular integral
\be
\langle A(\fglobal(\theta)) \rangle=\int_0^1 dp\ A(p)\ P(f(\theta)=p|\data)\ .
\ee

Particularly in cases of few experimental observations, the expected landscape $\langle \fglobal(\theta)\rangle$ does not necessarily approximate the exact landscape $F(\theta)$ well for any value of $\theta$,
but there can be substantial uncertainty. 
Since a strategy ignoring this will likely miss the global maximum, it is essential to take it into account, {\it e.g.}
in terms of the variance or higher order statistical moments.

A strategy based on both expectation and uncertainty allows 
to identify values of $\theta$ for which large values of $F$ 
can be expected.
In practice this can be realized in terms of the acquisition function
\be
a(\theta)=\langle F(\theta)\rangle+\alpha\sigma(\theta)\ ,
\label{eq:UCB}
\ee
with a real, non-negative scalar $\alpha$ that balances the weight of the expected landscape and the confidence in this estimate.
Using the value of $\theta$ that maximizes $a(\theta)$
for the next query to the experiment allows the algorithm to broadly explore the control landscape, thus minimizing the risk of getting trapped in a local extremum.

The choice of the numerical value of $\alpha$ can be made depending on the practical requirements.
A small value will result in rapid convergence to some reasonably good solution which might not be the best solution available, whereas a large value reduces the risk of not finding the best solution, but will typically result in slower convergence.
In practice, one may also change the value as the optimization progresses, starting with a large value favoring exploration of the entire landscape, followed by a decrease to focus on the particularly promising domains.

\subsection{Implementation}
\label{sec:integrals}

Sections \ref{sec:surmodel} to \ref{sec:decisionrules} describe conceptually the basic structure of \bo and its specificities in the presence of substantial measurement noise.
In this final subsection, we will discuss the most relevant aspects for the explicit implementation of these concepts,
but we invite any reader who is more interested in the performance of the present techniques than the underlying details to jump ahead to section \ref{sec:atwork} .

\subsubsection{The prior distribution}
\label{sec:gaussian}

In order to implement the algorithm laid out so far, one needs to define the prior distribution $P(f)$ of surrogate models explicitly.
Crucially, this does not require an explicit parametrization of possible control landscapes, but this can be done elegantly in terms of Gaussian processes as detailed in the following.
 
A random process extends the concept of a vector of random variables to an infinite collection of random variables.
Any function $g$ can thus be understood as the continuum limit of a set of discrete random numbers $g(\theta_i)$.
Since optimization landscapes are functions of continuous control parameters,
random processes are the appropriate mathematical structure for the present purposes.
In particular, the description of potential landscapes in terms of random processes avoids the necessity of an explicit parametrization,
but permits to ensure that properties like continuity and differentiability of the control landscapes are guaranteed.

A random process such that any finite collection of variables $[g(\theta_1), \hdots, g(\theta_N)]$ follows a Gaussian distribution, is referred to as a Gaussian process. 
While a Gaussian distribution is parametrized by a mean vector and a positive semi-definite covariance matrix, a \gp is specified by a mean function $\mu(\theta)=\langle g(\theta) \rangle$ and a positive semi-definite kernel function $k(\theta,\theta^\prime)$ which defines the covariance
\be
\langle g(\theta)g(\theta^\prime)\rangle-\langle g(\theta)\rangle\langle g(\theta^\prime)\rangle=k(\theta,\theta^\prime)\ ,
\ee
between $g(\theta)$ and $g(\theta')$ for arbitrary $\theta$ and $\theta'$, where the symbol $\langle\circ \rangle$ denotes the average, under this distribution.
This covariance is typically taken to be exponentially decaying in the distance between the two arguments,
which implies that knowledge of a function at some point $\theta$ allows one to estimate the function around this point with high confidence,
whereas the ability for prediction
decreases with the distance from this point.

A typical function $g(\theta)$ consistent with a Gaussian process will generally not qualify as proper surrogate model, since it will not be bound to the interval $[0,1]$ as a probability has to.
A Gaussian process on its own is therefore not suitable for the present purposes, but
one can arrive at a suitable distribution with a {\it squashing function} $\pi$,
such that the functions $f(\theta)=\pi(g(\theta))$ are bounded by the interval $[0,1]$.
With $g(\theta)$ resulting from a Gaussian process, this construction yields a probability distributions for functions $f(\theta)$ with the desired properties.
In the following we will use the cumulative distribution function of a standard normal distribution $\pi(x)=\int_{-\infty}^xdy\ \exp(-y^2/2)/\sqrt{2\pi}$,
but essentially any monotonic function mapping the real axis to the interval $[0,1]$ could be used.

With a specific choice of mean function $\mu(\theta)$ and kernel $k(\theta,\theta^\prime)$, one thus has the desired explicit realization of the prior distribution $P(f)$.
Without any prior knowledge about the optimization landscape one would typically choose a vanishing mean, $\mu(\theta)=0$ resulting in a flat expected landscape $\langle f\rangle=1/2$.

There are many possible choices for the kernel function $k(\theta,\theta^\prime)$~\cite{williams2006gaussian},
and we will choose them from the class of Mat\'ern functions
\be
C_{j}(\theta-\theta^\prime)=V\new{R_j(x)}\exp(-x)\ ,
\label{eq:kernel}
\ee
with $x=|\theta-\theta^\prime|/l$, defined in terms of a correlation length $l$,
and a variance $V$.
Depending on the optimization problem at hand,
we will use the three different polynomials
\bqa
\new{R_0(x)}&=&1\ ,\label{eq:kernela}\\
\new{R_1(x)}&=&1+\sqrt{3}x\ ,\label{eq:kernelb}\ \mbox{and}\\
\new{R_2(x)}&=&1+\sqrt{5}x+5/3x^2\ .\label{eq:kernelc}
\eqa
The choice of polynomials depends mostly on the expected roughness of the control landscape.
\new{$R_2$} effectively ensures that any $f$ that occurs with non-vanishing probability is at least twice differentiable.
\new{$R_1$} results in control landscapes that are at least once differentiable, and \new{$R_0$} enforces only that the landscapes are continuous~\cite{williams2006gaussian}.

Choosing suitable values for the variance $V$ and correlations length-scale $l$ would require some knowledge of the scales over which the value of $F$ changes.
Since even a very limited amount of data can be used to identify the values of $V$ and $l$, one can adapt them during the course of the \new{optimization}.
We will do so following standard practice \cite{williams2006gaussian,shahriari2015taking} by minimizing the log marginal likelihood term $\log P(\data)$, with $P(\data)$ defined in context of Eq.~\eqref{eq:post}.

\begin{figure*}[t!]
	\includegraphics[width=0.33\textwidth]{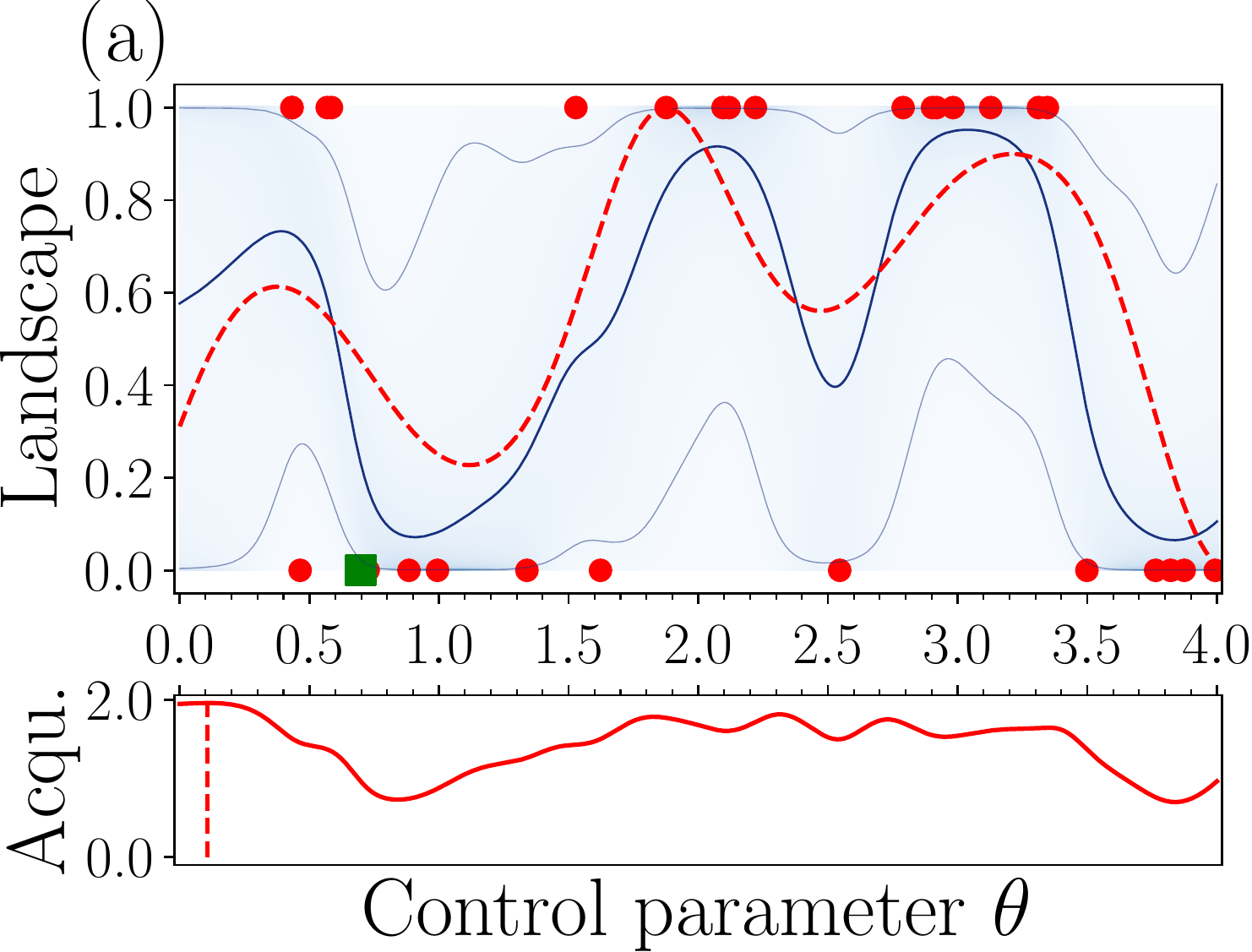}
	\includegraphics[width=0.31\textwidth]{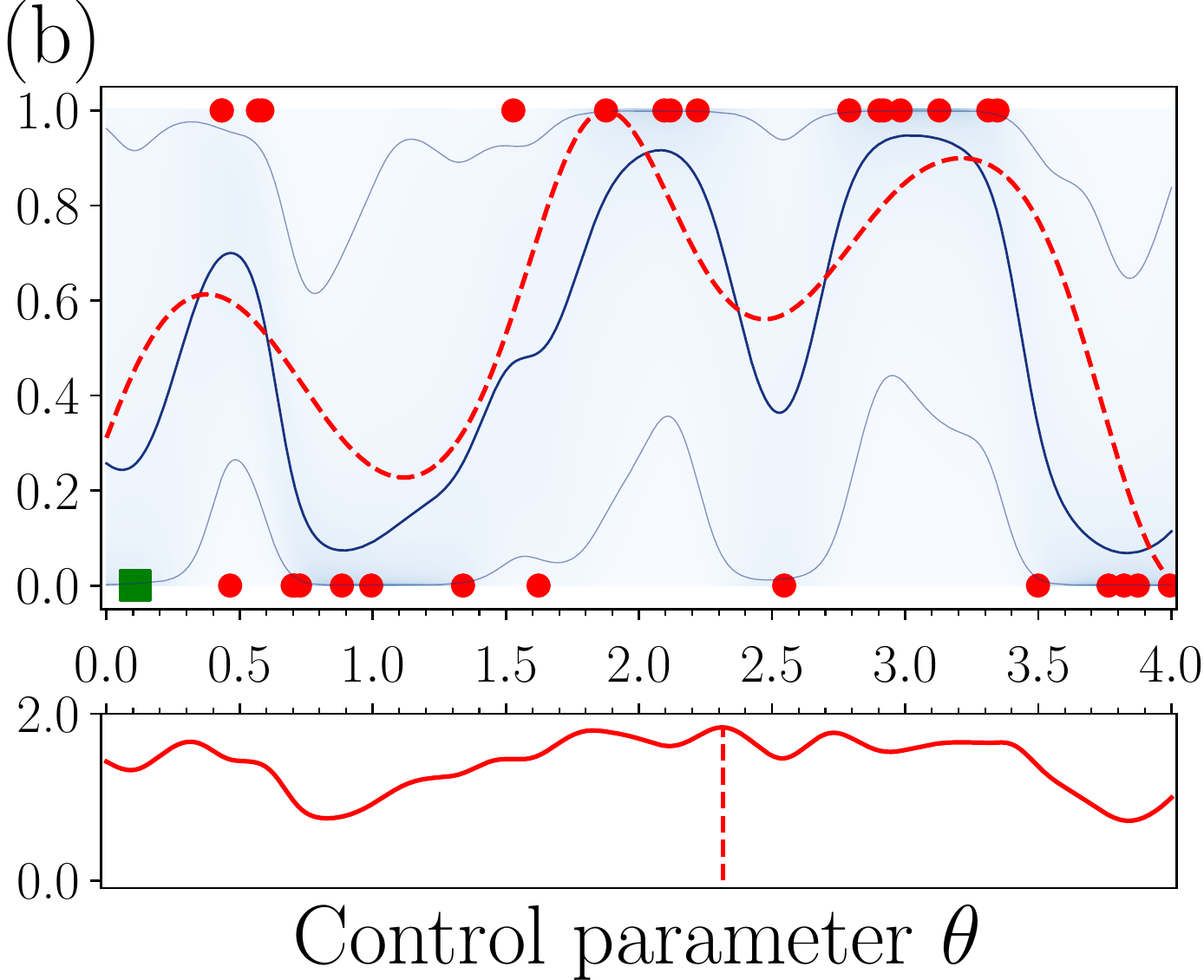}
	\includegraphics[width=0.31\textwidth]{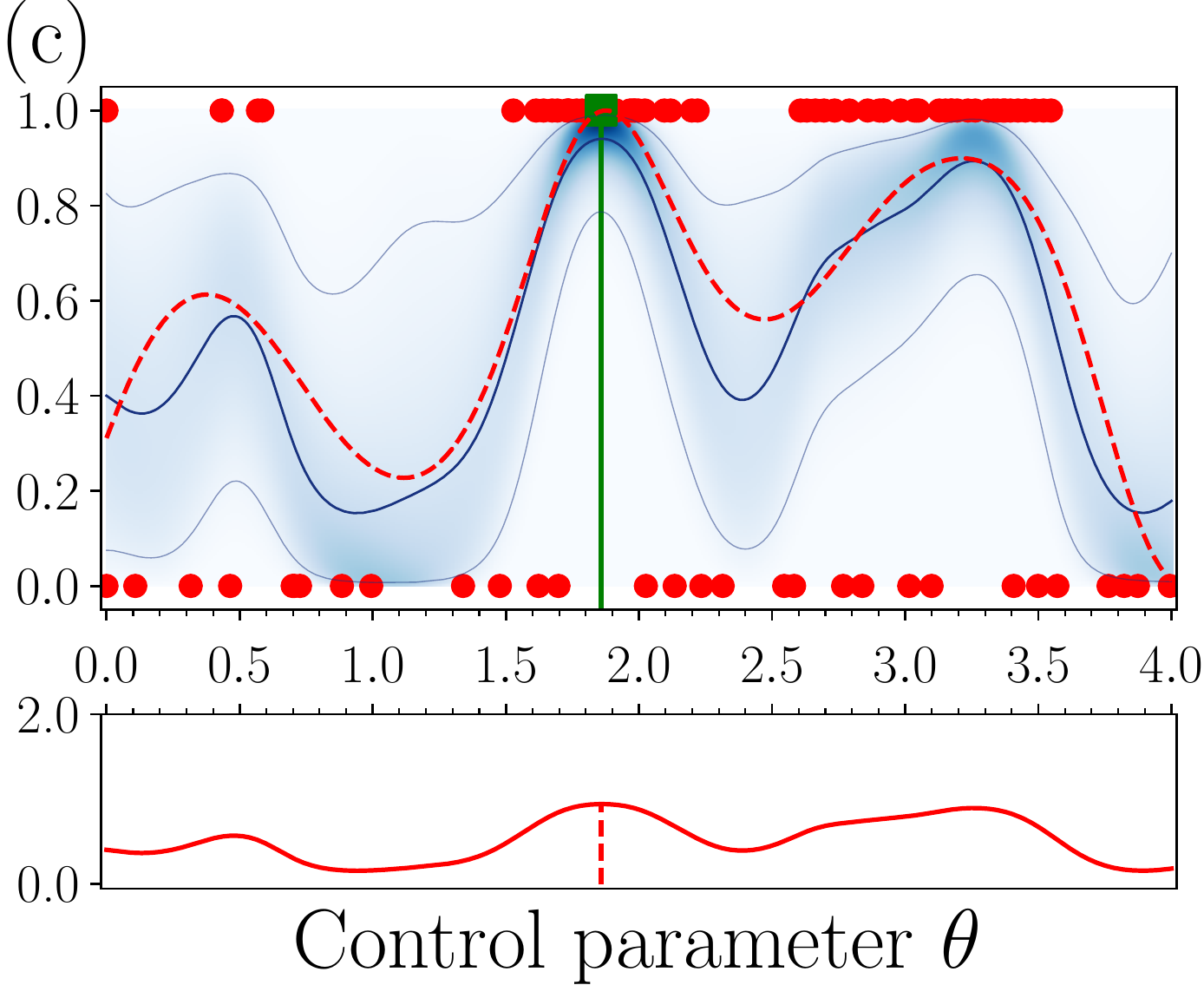}
	\caption{
Three stages of a typical run of Bayesian optimization with extremely bad measurement statistics, after $30$ (a) , $31$ (b) and $100$ (c) steps.
The top panel depicts the actual control landscape (dashed red), the expected control landscape (solid blue) and a $95 \%$ confidence interval (solid grey) around the expected landscape.
The actual probability density is represented by shades of blue.
Simulated, random outcomes of a projective measurement are depicted by red and green dots, where green corresponds to the most recent data point.
Assuming a sufficiently smooth landscape, it is possible to predict control landscapes ranging continuously between the values $0$ and $1$ despite the digital data.
The lower panel depicts the acquisition function (Eq.~\eqref{eq:UCB} with $\alpha=4$ for (a) and (b) and $\alpha=0$ for (c)),
the maximum of which determines the value of $\theta$ for the next upcoming step in the iterative algorithm.
During the course of the \new{optimization} (from left to right), the surrogate model becomes a better approximation of the actual landscape in the vicinity of the maxima, but the algorithm avoids the effort that would be required to approximate the landscape well in other domains.}
	\label{fig:bo}
\end{figure*}

\subsubsection{Integration}
The explicit construction of $P(f(\theta)|\data)$ in Eq.~\eqref{eq:predprob} requires the solution of an $\ns$-dimensional integral.
Already after a few steps in the iterative control algorithm a numerical evaluation of such an $\ns$-dimensional integral is prohibitively expensive.
An efficient, but still accurate estimate is thus essential for the practical value of the control scheme.

In the following we will discuss, how this is routinely done in many applications, and how the specificities of poor statistics require a different approach in order to maintain accuracy.

\paragraph{Integration with binomial noise model}
\label{sec:binomnoise}

Many problems of \bo  are based on entirely Gaussian models
whose integrals have analytic solutions.
In the present case, however,
only the first factor $P(f(\theta)|\vec f)$ in Eq.~\eqref{eq:predprob} is Gaussian, because of the underlying Gaussian process,
but the second factor $P(\vec f|\data)$, is non-Gaussian because $P(\data|\vec f)$ is given in terms of the binomial distribution in Eq.~\eqref{eq:PDf}.

In order to perform the required integration efficiently, we will employ the Laplace approximation \cite{williams2006gaussian} for $P(\data|\vec f)$ due to its conceptual simplicity.
It entails approximating $P(\data|\vec f)$ by a Gaussian distribution, such that $P(f(\theta)|\data)$ in Eq.~\eqref{eq:predprob} becomes an $\ns$-dimensional, analytically solvable integral over the product of two Gaussian distributions.
It is thus possible to efficiently perform all integrations despite the detailed and non-Gaussian underlying noise given in Eq.~\eqref{eq:PDf},
and we will discuss the implications of the approximate integration in detail later-on in Sec.~\ref{sec:ghz}.

\paragraph{Gaussian noise}
\label{sec:gaussiannoise}

In a standard approach that does not aim at detailed modeling of measurement noise,
one would estimate the probabilities $p_k$ of measurement results
in terms of the frequencies $p_k^{(e)}=n_k/N$ extracted from the observations.
These estimated probabilities admit an estimation of the figure of merit $F$ as
\be
F\simeq F_e={\cal Q}(p_1^{(e)}, \hdots, p_K^{(e)}),
\ee
with the function $\mathcal{Q}$ defined in Eq.~\eqref{eq:Fp}.

The fact that this estimate does not necessarily coincide with the exact value of $F$ is then taken into account with a phenomenological Gaussian probability distribution
\be
P(F_e(\theta)|f(\theta))=\frac{1}{\sqrt{2\pi}\sigma}\exp\left(-\frac{(F_e(\theta)-f(\theta))^2}{2\sigma^2}\right),
\label{eq:gauss}
\ee
which, in this case, is defined in terms of a single surrogate model $f$.
In contrast to the binomial modeling above, there is the phenomenological parameter $\sigma$;
its value is not determined by basic principles, but needs to be chosen in accordance with observations similarly to the choice of values for the \gp discussed above in Sec.~\ref{sec:gaussian}.

\section{
Poor statistics at work}
\label{sec:atwork}
In order to demonstrate the strength and limitations of the control algorithm developed in Sec.\ref{sec:algorithm}, we will apply it to four exemplary problems.

The problem of state preparation of a single qubit discussed in Sec.\ref{sec:singlequbit} is intended to be of pedagogical nature, giving insight into the workings of the methodology.
This is followed by the analysis
of the state-preparation of a three-qubit GHZ state in terms of single- and two-qubit gates (Sec.\ref{sec:ghz}),
and the preparation of a Mott-insulating quantum phase in a Bose-Hubbard system (Sec.\ref{sec:1dbosons}).
\new{In addition to these demonstrations based on numerical simulations,
the problem of state preparation on a publicly available NISQ (Noisy Intermediate-Scale Quantum) device~\cite{ibmq} discussed in Sec.~\ref{sec:ibmchips} gives experimental evidence of the benefits of the present approach.}

In all these examples the efficiency of the present framework will be judged in terms of its ability to converge towards good solutions with minimal experimental effort.
This effort is quantified as the total number $\nq$ of experimental runs,
{\it i.e.} individual cycles of preparation, evolution and measurements.
Assessing the expectation value of a single observable with $\nr$ repetitions, for example, requires $\nq=\nr$ runs,
whereas assessing several commuting observables in terms of a single-shot measurement requires only a single run. 
\new{The explicit examples in this section will demonstrate,
that the present framework permits to reduce the number of experimental runs by working with few repetitions $\nr$, {\it i.e.} poorly resolved expectation values of observables, allowing to explore more control pulses at given number $\nq$ of experimental runs.
If a given experiment poses a limitation on the number of control pulses to be tested, one may select the lowest number $\nr$ of repetitions that is sufficient to converge within this number of pulses.}

\new{With the exception of the illustrative example of Sec.~\ref{sec:singlequbit},
the present framework with proper binomial modeling of measurement noise (as developed in Sec.~\ref{sec:algorithm}) will be systematically benchmarked against 
\bo with Gaussian modeling of measurement noise (as sketched in Sec.~\ref{sec:gaussiannoise}),
and the Simultaneous Perturbation Stochastic Approximation (SPSA) framework~\cite{spall1992multivariate}. 
These two techniques are especially competitive when data are noisy and limited~\cite{wigley2016fast,2017arXiv171205771O,Henson2018,zhu2018training,nakamura2019non,PhysRevA.91.052306,Kandala2017,PhysRevLett.117.040402,Havlicek2019}
and as such provide strong benchmarks to compete with.}

\new{As the success of any optimization depends on the random choice of initial guesses but also
on measurement data which are intrinsically probabilistic,
most results presented in Sec.~\ref{sec:atwork} correspond to optimizations that are repeated several times, with the exception of Fig.~\ref{fig:bo} in Sec.~\ref{sec:singlequbit} which depicts one specific instance of an optimization for illustrative purposes. 
For results obtained based on numerical simulations, these repetitions are taken over different random seeds, and measurement data are sampled following Born's rule.}

All the following \new{numerical} simulations and optimizations are based on
QuTip~\cite{JOHANSSON20131234}, QuSpin~\cite{SciPostPhys.2.1.003}, GPy~\cite{gpy}, and GPyOpt~\cite{GPyOpt}\new{; and interfacing with the 
NISQ device in Sec.~\ref{sec:ibmchips} is based on
Qiskit~\cite{qiskit}.}

\subsection{A single qubit to warm up}
\label{sec:singlequbit}

A simple toy model of pedagogical value is given by the state $\ket{\psi(\theta)}$
of a single qubit parametrized with a scalar control parameter $\theta$.
The goal lies in maximizing the fidelity with respect to a given state $\ket{\phi}$, and we assume for simplicity that a projective measurement in a basis including the state $\ket{\phi}$ can be taken.
In the present case, the target functional is thus directly the probability to project onto $\ket{\phi}$, {\it i.e.} $F(\theta)= |\langle \phi \ket{\psi(\theta)}|^2$.

Rather than defining a control landscape in terms of the dynamics induced by a specific Hamiltonian, we will consider the exemplary function
\be
F(\theta)=\sin^2 \Big( \frac{1}{2}\sin\Big(3\theta+\frac{9}{10}\Big)+ \frac{3\theta}{2}+\frac{9}{20}\Big)\ .
\ee
This landscape, depicted by a dashed red line in Fig.~\ref{fig:bo}, reaches the maximum value of $1$ and has two additional local maxima in the interval $[0,4]$;
it is thus well suited to illustrate the ability of the algorithm to distinguish local from global extrema.

Fig.~\ref{fig:bo} depicts the progress in data acquisition and the estimates of the control landscape during the course of the optimization using the kernel defined in Eq.~\eqref{eq:kernel} with $P_{2}$ given in Eq.~\eqref{eq:kernelc}.
The  optimization starts after taking $30$ initial single-shot projective measurements, for randomly chosen $\theta_j$ ($j=1,\hdots,30$) as shown in Fig.~\ref{fig:bo} (a).
The measurement results depicted by red dots and a green square can only adopt the values of $0$ and $1$, and each value is randomly drawn following the actual control landscape. 
Based on those $30$ observations, one obtains a rough estimate of the actual control landscape; the expected landscape $\langle f(\theta)\rangle$ is depicted in blue, a $95\%$ confidence interval is depicted by the grey contour, and the actual probability density is represented by shades of blue.
As one can see the expected control landscape captures the most salient features like the approximate locations of the three maxima, but, given the very limited data it fails to reproduce accurately the actual landscape.
In particular, it fails to distinguish the local from the global maximum.

Based on this model the decision rule described in Sec.~\ref{sec:decisionrules} determines for which value of $\theta$ to take the next measurement. The acquisition function (Eq.~\eqref{eq:UCB}), with a value of $\alpha=4$, is depicted in red at the bottom and takes its maximal value for $\theta_{31}\approx0.1$ (red vertical line). This decision reflects the high uncertainty in the model in this region where no experimental observations are available yet.

Fig.~\ref{fig:bo} (b) contains the additional data point (square green) resulting from the projective measurement with the value $\theta_{31}$ of the control parameter.
The updated surrogate model in (b) shows that the surrogate model in (a) has over-estimated the value of $F(\theta_{31})$.
In addition to the better estimate of this value, the uncertainty around $\theta_{31}$ has also slightly decreased from (a) to (b).
As a result of these two effects the next probe is taken for the value $\theta_{32}\approx2.3$, {\it i.e.} in the vicinity of the maximum where the value of expected control landscape is moderately high and, in addition, the uncertainty is large.

Fig.~\ref{fig:bo} (c) depicts the surrogate model after $100$ queries to the experiment.
As one can see, the framework suggested several probes in the vicinity of the three maxima, and managed to avoid queries that would not have resulted in substantial added value.
Consequently, the surrogate model approximates the actual control landscape more accurately close to these maxima.

After these $100$ iterations the algorithm has identified the value $\theta_f=1.86$ (as indicated in green in Fig.~\ref{fig:bo} (c)) as optimal, which is very close to the true optimal value $\theta_o=1.876$.
With the value $F(\theta_f)=0.996$ of the actual control landscape,
the algorithm has thus found a solution with an infidelity of $0.004$,
after a number of measurements that would have only been enough to determine a single point of the control landscape with a resolution of $0.01$.

\subsection{GHZ states}
\label{sec:ghz}

\begin{figure}
	\includegraphics[width=0.49\textwidth]{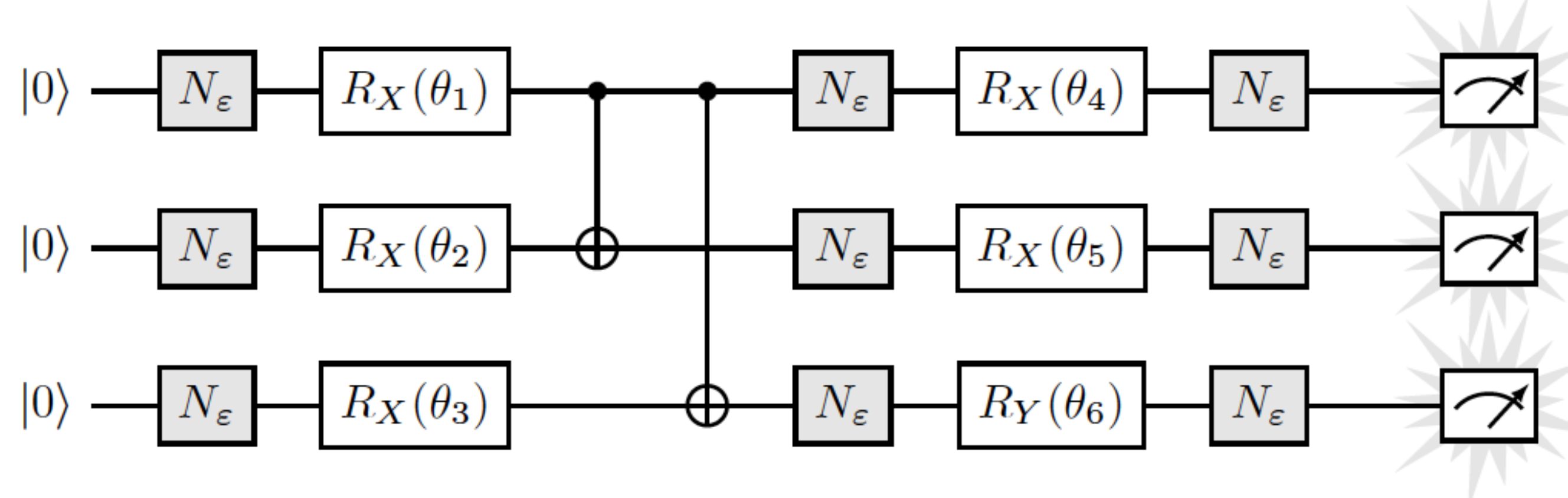}
	\caption{
	Gate sequence for the preparation of a three-qubit GHZ state in terms of two controlled-not gates and $6$ single qubit gates with adjustable parameter $\theta_j$.
	The addition of noisy unitaries $N_\varepsilon$ and readout error helps to demonstrate optimal control in the presence of experimental noise in addition to the measurement shot noise.}
	\label{fig:circuit}
\end{figure}

With the flavor of the workings of optimal quantum control based on poor statistics laid out in Sec.~\ref{sec:singlequbit},
we can now proceed to a more challenging control problem.
We consider the gate sequence depicted in Fig.~\ref{fig:circuit} for the preparation of a three-qubit GHZ state $\ket{\Psi}=(\ket{000}+\ket{111})/\sqrt{2}$.
that could be realized for example in trapped ion \cite{Roos1478,Leibfried1476,Wright2019} or superconducting qubit \cite{DiCarlo2010,Neeley2010,Arute2019} hardware.
It consists of five single qubit gates $R_{x}(\theta_j)=\exp(i\theta_j\sigma_{x})$ ($j=1,\hdots,5$), one single qubit gate $R_{y}(\theta_6)=\exp(i\theta_6\sigma_{y})$, and two C-NOT gates as shown in Fig.~\ref{fig:circuit}.

The goal is to find suitable angles $\theta_j \in [0,2\pi]$ such that the circuit maps the initial state $\ket{000}$ into the GHZ state $\ket{\Psi}$.
Later-on, we will also consider additional noisy unitaries $N_\varepsilon$, but for the moment, they are treated as identities.

\subsubsection{Control targets} 
The fidelity $F$ for the state $\varrho$ resulting from the gate sequence is defined as $F=\matel{\Psi}{\varrho}{\Psi}$.
Unless one is able to perform a projective measurement in a basis including the state $\ket{\Psi}$, however, one is bound to perform measurements on each individual qubit, as indicated by the three detectors in Fig.~\ref{fig:circuit}.
In practice, one would therefore construct the fidelity in terms of local measurements.

Fidelities are often expressed in terms of expectation values of observables.
In the case of a GHZ state, it can be decomposed as
\be
F=\frac{1}{8}\Bigl(1+\sum_{k=1}^4\mbox{tr}(\varrho S_k)-\sum_{k=5}^7\mbox{tr}(\varrho S_k)\Bigr),
\label{eq:fid0}
\ee
where
$S_1=\sigma_x\otimes\sigma_x\otimes\sigma_x$,
$S_k$ with $k=2,3$ and $4$ are the permutations of $\um\otimes\sigma_z\otimes\sigma_z$,
and 
$S_k$ with $k=5,6$ and $7$ 
are the permutations of $\sigma_x\otimes\sigma_y\otimes\sigma_y$.

Since the framework developed so far allows us to estimate probabilities rather than expectation values, it is more appropriate to express the fidelity in Eq.~\eqref{eq:fid0} in terms of probabilities for measurement outcomes.
Denoting $P_k=(S_k+\um)/2$ as the projector onto the subspace spanned by the eigenstates with eigenvalue $+1$ of the observable $S_k$,
the state fidelity for the GHZ state can be written as
\be
F=\frac{1}{4}\Bigl(\sum_{k=1}^{4}\mbox{tr}(\varrho P_{k})-\sum_{k=5}^{7}\mbox{tr}(\varrho P_{k})\Bigr).
\label{eq:fid}
\ee

In contrast to the simpler case discussed above in Sec.~\ref{sec:singlequbit}, the fidelity is thus not given by the probability of one single measurement outcome,
but it is given in terms of the seven probabilities $p_k=\mbox{tr}(\varrho P_{k})$.
It is thus necessary to employ seven surrogate models $f_k$ as detailed in Sec.~\ref{sec:decisionrules},
and to estimate the state fidelity as
\be
\fglobal =\frac{1}{4}\left(\sum_{k=1}^{4}f_k-\sum_{k=5}^{7}f_k\right),
\ee
consistently with Eq.~\eqref{eq:fid}.

\subsubsection{Optimization}
\label{sec:optimization}

\begin{figure}
	\includegraphics[width=0.49\textwidth]{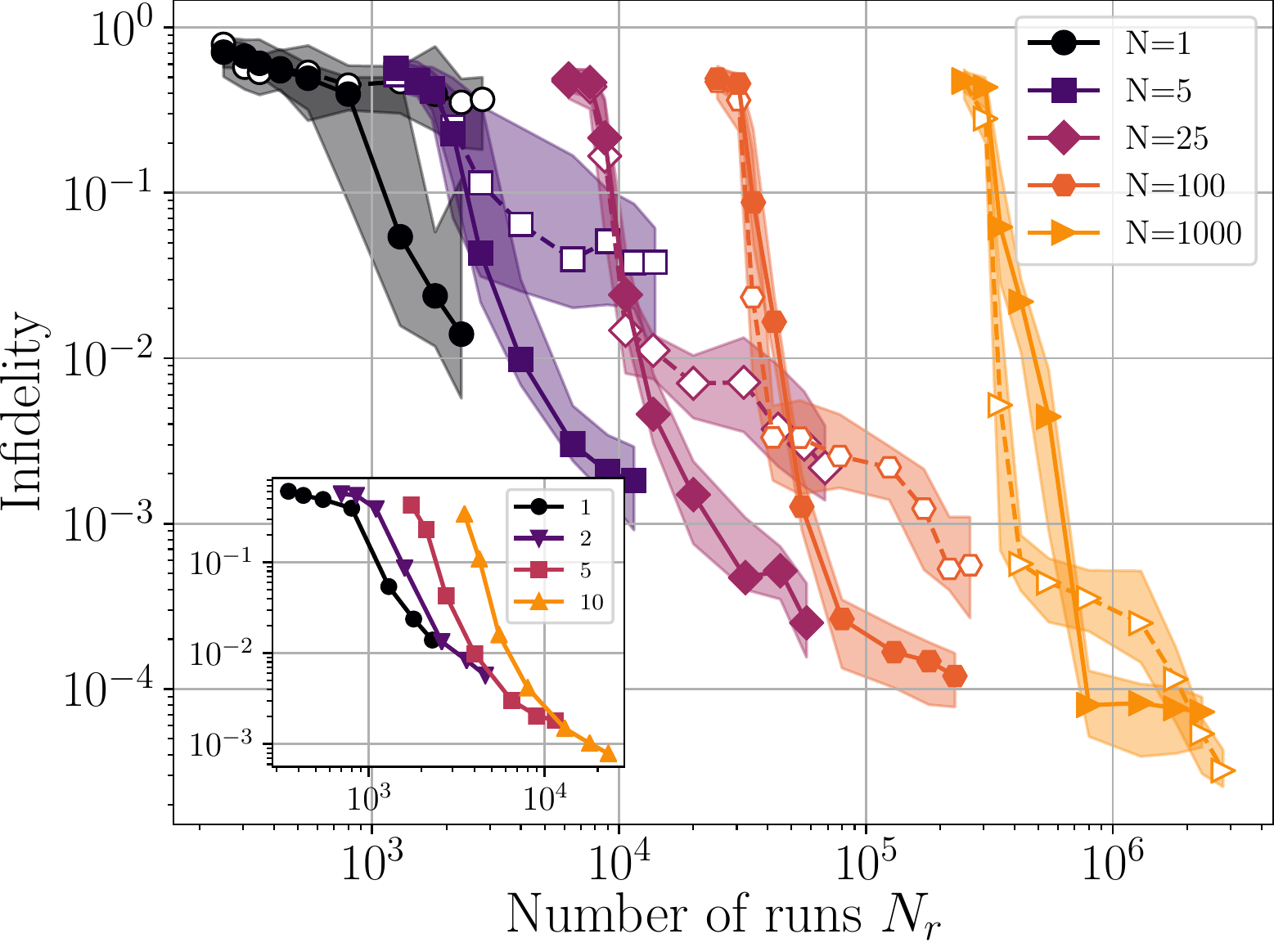}
	\caption{
Convergence of the control algorithms towards low infidelities $\mathcal{I}$ as a function of the total number $\nq$ of runs of the circuit required.
The median values of $\cal{I}$ are indicated by solid symbols for the binomial modeling of measurement noise and by empty symbols for Gaussian modeling.
The different colors correspond to different numbers of repetitions of each measurement, ranging from $\nr=1$ in black to $\nr=\num{1000}$ in orange.
Interquartile intervals are depicted by shaded regions.
Optimizations with the proper binomial modeling converge substantially faster than optimizations with Gaussian modeling of measurement noise, 
except in the case with large number of repetitions $\nr=\num{1000}$.
\new{The inset depicts results for
fewer repetitions ranging from $\nr=1$ in black to $\nr=10$ in orange, and binomial modeling only. Interquartile intervals are not displayed, for visual clarity.}  
}	
\label{fig:binvsgauss}
\end{figure}

\new{The following optimizations are obtained with $R_0$ (Eq.\eqref{eq:kernela}) for binomial modeling and with $R_2$ (Eq.\eqref{eq:kernelc}) for Gaussian modeling.}
In all these examples, we will consider convergence in terms of the total number $\nq$ of runs required to reach a given fidelity.
Since the three observables $S_5$, $S_6$ and $S_7$ commute, an optimization based on the assessment of the fidelity (Eq.~\eqref{eq:fid}) for $M$ different sets of control parameters,
with each measurement repeated $\nr$ times, requires $\nq=5M\nr$ runs.

With few repetitions for a measurement, in particular, in cases of single-shot measurements ($\nr=1$), any individual run of an optimization has a substantial probabilistic component.
The following comparisons, will thus be based on statistics of the infidelity ${\cal I} = 1 - F$, such as the median and quartiles, over $30$ random instances of the same control task.

The main frame of Fig.~\ref{fig:binvsgauss} depicts the decrease in infidelity as a function of the total number of runs $\nq$.
Different colors correspond to different numbers of repetitions ranging from the single-shot case $\nr=1$ to $\nr=\num{1000}$; results from binomial and Gaussian modeling are depicted with filled and empty symbols.
Fluctuations in convergence towards high fidelities due to statistical fluctuations in measurement results are characterized in terms of the interquartile intervals depicted by shaded regions.

In the case of single-shot measurements (depicted in black), control based on Gaussian modeling essentially fails,
whereas binomial modeling yields infidelities in the percent regime.
With larger number of repetitions $\nr$, also Gaussian modeling results in low infidelities, but for any given choice of $\nr$ (with the exception of $\nr=1000$), control based on binomial modeling converges faster;
after a given number of runs $\nq$, it yields infidelities about an order of magnitude lower than control based on Gaussian modeling.
The gap between the two methods decreases slowly as $\nr$ increases,
and for $\nr = \num{1000}$ one might have expected Gaussian modeling to become comparable to binomial modeling.
There are, however, instances in which the Gaussian version slightly outperforms the binomial version. 
This can be attributed to the Laplace approximation (Sec. ~\ref{sec:inpractice}).
Close to the extreme values of $0$ and $1$ of the underlying probabilities this approximation becomes less accurate \cite{williams2006gaussian} resulting in an over(under)-estimation of the actual value of the probabilities.
Signatures of this effect can also be seen in Fig. ~\ref{fig:bo}(c), where the expected control landscape is systematically below the true landscape in the vicinity of the global maximum.

Even though this effect can be reduced at the expense of higher computational effort \cite{nickisch2008approximations}, this does not seem necessary, since these stagnations occur in a regime in which the Gaussian modeling becomes appropriate and analytic solutions for integrals are available (Sec.~\ref{sec:gaussiannoise}).

One can see in Fig.~\ref{fig:binvsgauss} -- especially in the inset --  that in all cases where there is data for different values of $\nr$ for a given infidelity,
the optimization with the fewest repetitions performs best.
It thus seems to be always preferable to explore many points of the control landscape with few repetitions rather than trying to resolve the control landscape accurately for few points.
Reducing the number of repetitions of the same measurement, however, also implies additional computational cost:
the algorithm scales as $\mathcal{O}(\ns^3)$ \cite{williams2006gaussian} in terms of the number $\ns$ of iterations, and at constant number $\nq\propto \ns\nr$ of measurements a reduction of $\nr$ implies an increase in $\ns$.
That is, whereas fundamentally optimizations with the fewest repetitions seem to perform best, it can become necessary in practice to increase the number $\nr$ of repetitions.

\begin{figure}
\includegraphics[width=0.493\textwidth]{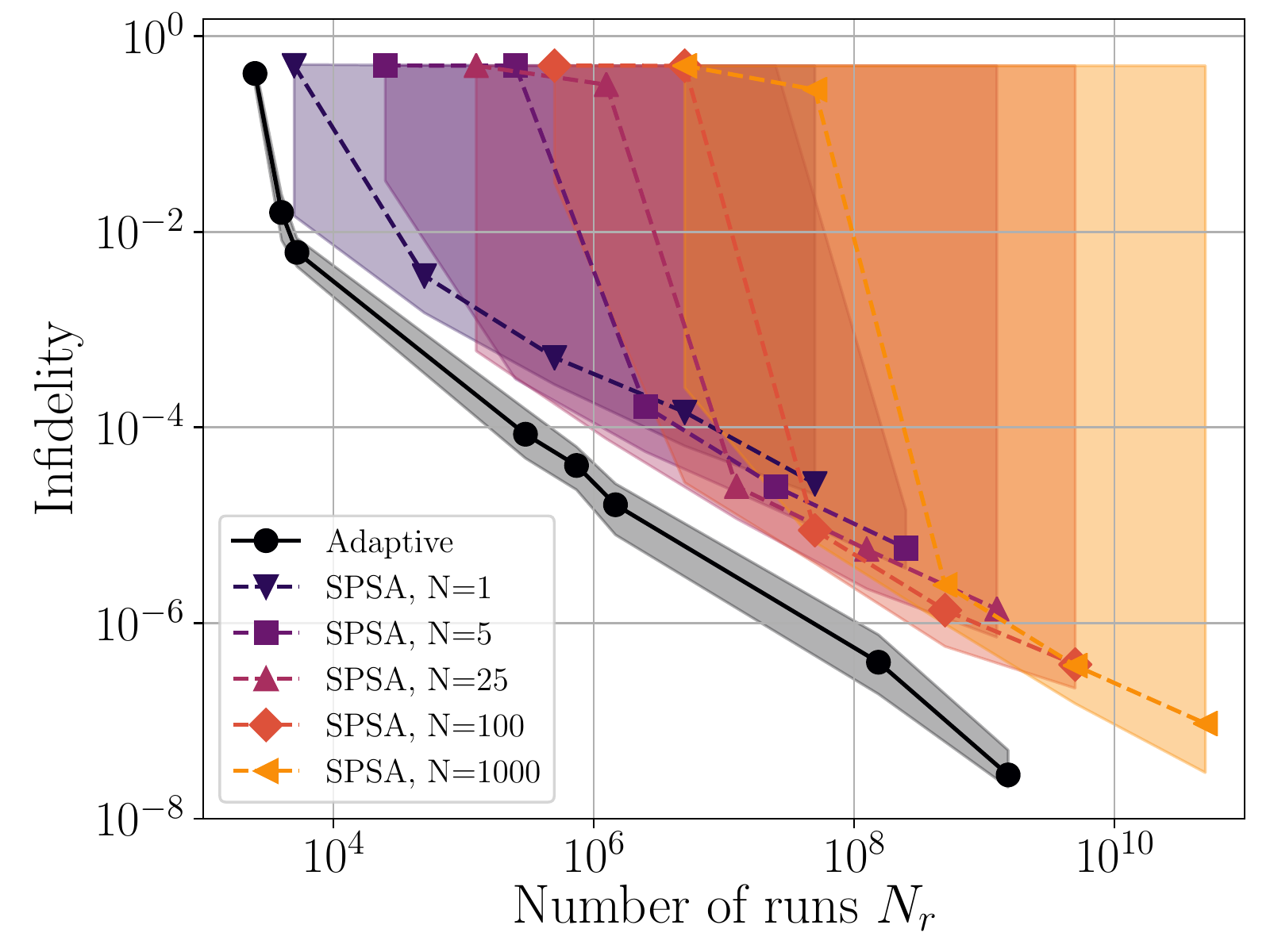}
\caption{Comparison of the adaptive strategy (black) explained in Sec.~\ref{sec:scalable} with optimizations performed using SPSA with a fixed number $\nr$ of \new{repetitions of each measurement} 
given in legend.
The adaptive method converges systematically faster than any of the other cases.}
\label{fig:adaptive}
\end{figure}

\subsubsection{Adaptive strategy}
\label{sec:scalable}

Given that working with few repetitions of the same measurement ({\it i.e.} small values of $\nr$) is beneficial for fast convergence, but that this expensive benefit tails off at high fidelities, 
\new{one may use a variable number of $\nr$ for an easily implementable reduction of computational effort.}
We found it most practical to start the optimization with few-shot measurements ({\it e.g.} $\nr\leq10$) in order to rapidly identify good domains in parameter space.
As the search circles in on a good solution, one can restrict the exploration to a smaller domain;
dropping the data outside this domain reduces the numerical effort, and the number of repetitions $\nr$ can be increased as the search approaches a high-fidelity solution.

The black curve in Fig.~\ref{fig:adaptive} depicts the convergence of this strategy.
The initial $150$ steps of the optimization are performed with $\nr=5$ repetitions and binomial modeling of the noise, corresponding to the first three data points of the curve.
After those initial steps the parameter space is reduced around the $75$ best parameters probed so far, and the number of repetitions is increased.
The next three points, with $\nq$ close to $10^6$ runs of the circuit correspond to an additional $500$ steps in the optimization with Gaussian modeling and an increased number of $\nr=100$, $250$ and $500$ repetitions. Finally the two remaining points of the curve, with $\nq>10^8$ runs, correspond to an additional stage of optimization with Gaussian modeling starting with the solutions obtained with $\nr=500$ repetitions.
For this round, the parameter space is reduced once more, and $\nr$ is further increased to the values of $\nr=\num{50000}$ and $\nr=\num{500000}$ respectively.
Overall the adaptive strategy quickly locates good control parameters (infidelities below $1\%$ after $\num{5000}$ runs) and subsequently is able to refine them to reach infidelities as low as $\mathcal{I}\approx10^{-8}$.

The remaining five data sets (shown in color in Fig.~\ref{fig:adaptive}) correspond to optimizations performed using SPSA with a fixed number of repetitions $\nr$.
Although SPSA also shows convergence towards low infidelities, this convergence is substantially slower than with Bayesian optimization:
comparison of Fig.~\ref{fig:binvsgauss} and Fig.~\ref{fig:adaptive} for any given number of repetitions $\nr$ indicates dramatically faster convergence of the optimizations with binomial modeling than with SPSA, 
\new{both during the early stage of the optimization when the landscape is being explored broadly and later-on when solutions are being refined close to the optimum.}
Also the adaptive framework that enables Bayesian optimization with many runs $\nq$
clearly outperforms SPSA at any given number $\nq$ of runs.

In addition to that, the \bo with binomial modeling
is also rather robust to statistical noise in the measurement results,
whereas the interquartile range (shaded regions of the figure), shows that optimizations based on SPSA are subject to very large fluctuations in performance.
In fact, a significant fraction (between $25\%$ and $50\%$) of all optimizations does not get close to the achievable fidelities,
but gets trapped in local minima with infidelity of about $50\%$.

\begin{figure} 
	\includegraphics[width=0.49\textwidth]{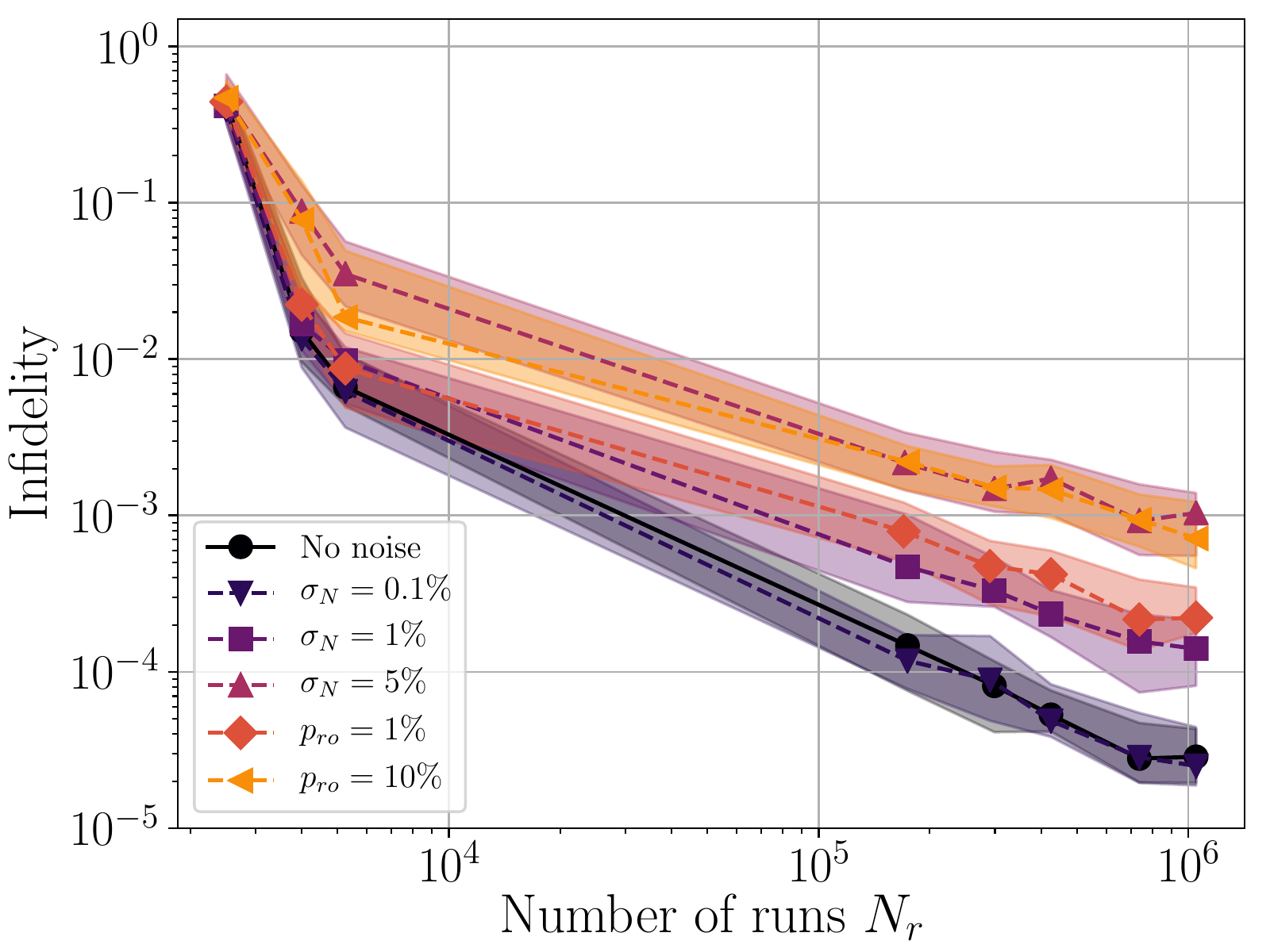}
	\caption{
	Convergence to low infidelities in the case of noisy circuits.
	The different colors correspond to different noise levels as specified in the legend 
	where $\sigma_N$ denotes the amplitude of the noisy unitaries and $p_{ro}$ the probability of a readout error.
	The noiseless case (black curve) is provided as a reference.}
	\label{fig:ghzadaptivenoise}
\end{figure}

\subsubsection{Additional noise}

Last but not least, we can demonstrate that the present control algorithm is not just able to cope with measurement noise, but that it can also be used to find good controls in the presence of additional sources of noise.
To this end, we can consider the noisy unitaries $N_{\varepsilon}=\sqrt{1-\varepsilon^2}\um+\varepsilon\vec n\vec\sigma$,
where $\vec\sigma$ is the vector of the three Pauli matrices.

In each individual run of the gate sequence, $\varepsilon$ is drawn from a Gaussian distribution with vanishing mean and width $\sigma_N$, and the vector $\vec n$ of unit length is drawn from a uniform distribution following the Haar measure.
In addition, we consider a finite probability of readout error $p_{ro}$
for each of the three detectors to yield the wrong result, {\it i.e.} `$0$' instead of `$1$' and vice versa.

In the presence of this noise with no correlations between the different noisy elements or between consecutive runs, it is fundamentally impossible to reach perfect fidelities.
The present goal therefore can not be to aim at unit fidelity in a noisy circuit, but to demonstrate the ability of \bo to find good solutions with a noisy circuit.
The infidelities in Fig.~\ref{fig:ghzadaptivenoise} therefore correspond to what is achieved with a noiseless circuit but for a control solution that has been optimized in the presence of noise.

Similarly to the other figures, Fig.~\ref{fig:ghzadaptivenoise} depicts a decrease in the median infidelity and quartiles as a function of the total number of runs of the circuit, for different values of $\sigma_N$ and $p_{ro}$, using the adaptive strategy described in the previous section with an initial number of $\nr=5$ repetitions.

Even though the control algorithm is entirely agnostic to the nature of the additional noise, there is rapid progress towards high fidelities.
Most cases feature a rapid drop in infidelity after a few thousand runs, with the $\sigma_N=5\%$ and $p_{ro}=10\%$ cases being the exception.
Even in the cases of very substantial additional noise, however, the optimization manages to converge well to low infidelities.

\begin{figure}
	\includegraphics[width=0.49\textwidth]{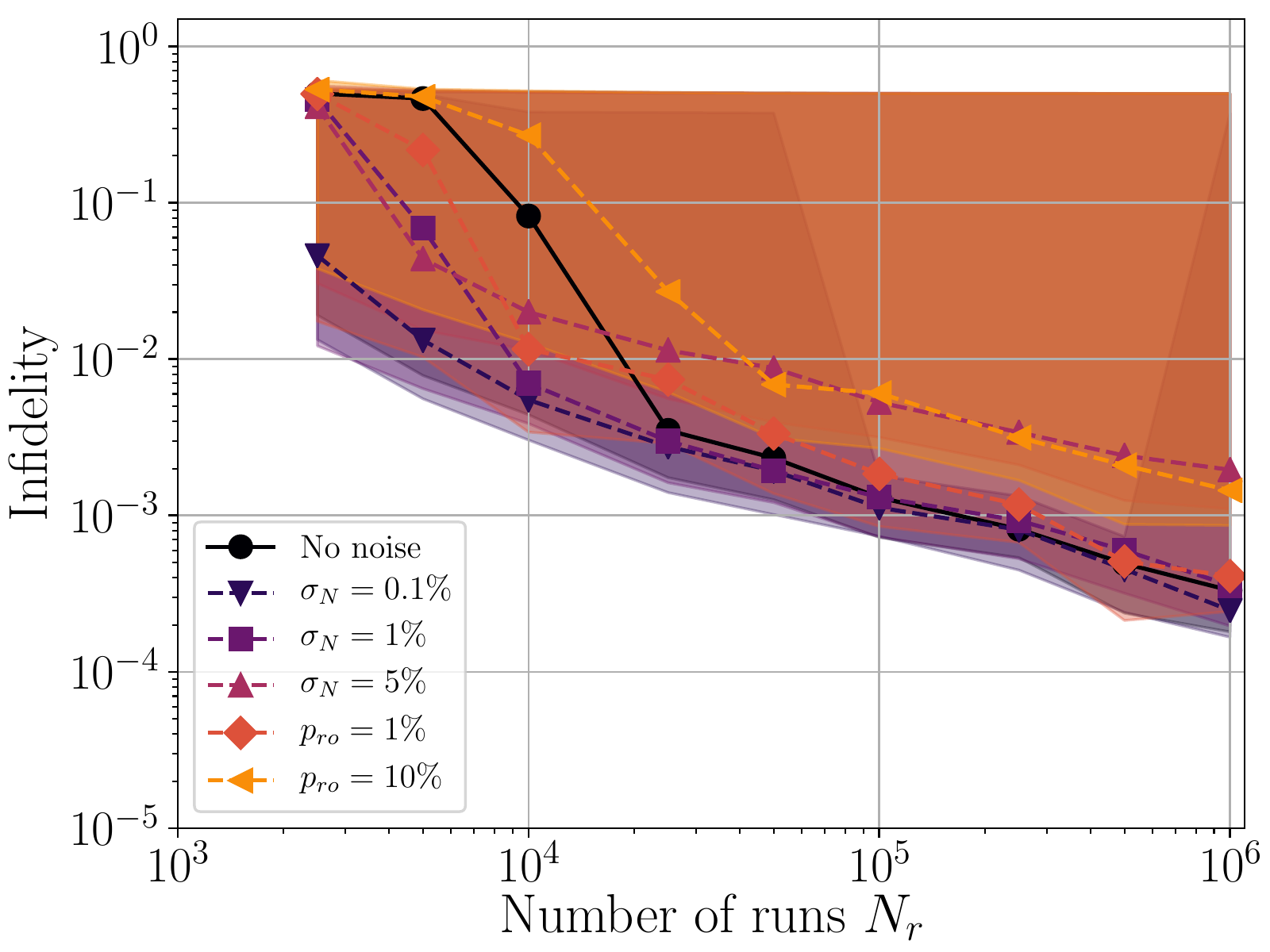}
	\caption{
\new{Convergence of SPSA in the presence of noise. All the results are obtained with single-shot measurements ($N=1$), since this gives the best results in the noiseless case.} 
}
	\label{fig:ghzspsanoise}
\end{figure}

\new{
These results can be compared to the convergence of the SPSA algorithm in the presence of noise
shown in Fig.~\ref{fig:ghzspsanoise} for the single-shot ({\it i.e.} $N=1$) measurements case.
Remarkably, noise can help in the convergence of the algorithm, especially for a low total number of runs ($\nq<\num{10000}$). 
The large interquartile intervals depicted in Fig.~\ref{fig:ghzspsanoise}, however, indicate that more than $25\%$ of the optimizations ended up trapped in local minima with high infidelity ($\mathcal{I}\simeq 50\%$),
whereas such unsuccessful instances of an optimization with binomial \bo are very rare.

For small levels of unitary noise ($\sigma_N = 0.1\%$) a typical instance of an SPSA optimization converges an order of magnitude slower than with binomial Bayesian optimization.
For higher noise levels this advantage is slightly reduced as statistical noise is not the main source of error anymore. 
Still, in all cases binomial \bo converges a factor of $2$ to $5$ faster than SPSA.
}

\subsection{Quantum many-body states}\label{sec:1dbosons}

Having analyzed the performance of the present approach with the example of a GHZ state, we can show with the example of a Bose-Hubbard system, that the findings in Sec.~\ref{sec:ghz}, are not specific to systems of qubits but that similar performance can be expected for control of a broad range of quantum systems.

\subsubsection{Model}

To this end, we consider a system of $\nbosons$ bosons in a one-dimensional arrangement with $\nsites$ sites and periodic boundary conditions.
The bosonic creation and annihilation operator at site $i$ of the chain is denoted $\hat{b}^{\dagger}_i$ and $\hat{b}_{i}$, and the local number operator reads $\hat{n}_i=\hat{b}^{\dagger}_i\hat{b}_{i}$.
The physics is described by the Bose-Hubbard model

\be
\label{eq:hambosons}
\new{H(t)=-J(t)\hat{T}+ U(t)\hat{V}\ ,}
\ee
with tunneling
\be
\hat{T}=\sum_i \hat{b}_{i+1}^{\dagger} \hat{b}_{i}+\hat{b}_{i-1}^{\dagger} \hat{b}_{i}\ ,
\ee
on-site interaction
\be
\hat{V}=\frac{1}{2}\sum_i \hat{n}_i(\hat{n}_i-1)\ ,
\ee
\new{as well as tunneling rate $J(t)$ and interaction strength $U(t)$ that can be tuned in actual optical lattice experiments~\cite{Greiner2002,Schreiber842}.
We will consider the transition from a non-interacting system
($U/J \rightarrow 0$) with the superfluid ground state as initial state to a Mott-insulating phase with $J/U \rightarrow 0$},
since the preparation of many-body quantum states with well-defined particle number per site is a 
prerequisite in many applications with neutral atoms~\cite{Labuhn2016,2017weiss,Bernien2017,PhysRevX.8.011032}.

\new{In order to account for the fact that the experimental values of tunneling rate and interaction strength are bounded,
while perfect, adiabatic protocols in finite time are obtained in the limit $J,U\to\infty$, we will parametrize $J(t)$ and $U(t)$ as
\bqa
J(t)&=&1-\Gamma(t)\ ,\mbox{and}\\
U(t)&=&\Gamma(t)\ ,\mbox{with}
\eqa
$\Gamma(t=0)=0$ and $\Gamma(t=T)=1$.}

\new{The goal of finding a time-dependent control function $\Gamma(t)$ such that the system ends up in the Mott-insulating ground state is getting increasingly difficult with decreasing sweep time $T$,
and the quantum speed limit \cite{van2016optimal} asserts that it
can not be achieved if $T$ is below the threshold of $\pi/\Delta$ with the minimal lowest energy gap $\Delta$ of the system Hamiltonian during the sweep.
In the following, we will choose the specific value of $T=3/2\ \pi/\Delta$,
which is substantially shorter than the time-scale required for an adiabatic transition.
An optimal solution under this time constraint will typically not be a slight deformation of a solution that is optimal close to the adiabatic regime.
Still this duration is
sufficiently long to find good control pulse $\Gamma(t)$ within the additional restrictions of an explicit parametrization.}

Due to the rapid growth of Hilbert space dimensions that limits the accessible range of exact diagonalization the following example is limited to a system with $\nsites=5$ sites and $\nbosons=5$ bosonic particles \new{which corresponds to a $16$-dimensional Hilbert space,
taking into account selection rules resulting from the system's translational and inversion symmetries.}

\subsubsection{Control function}

To be amendable to optimization the time-dependent control $\Gamma(t)$ is parametrized in terms of $\nparams$ parameters $\theta_i$ corresponding to points equally spaced in time, {\it i.e.}
\be
\Gamma(t_i) = \theta_i\ ,\hspace{.5cm}\mbox{with}\hspace{.5cm}t_i=\frac{i}{\nparams+1}T\ .
\ee
The parameters $\theta_i$ for $i=1,\hdots,\nparams$ are bound to the interval $[0,1]$,
and the values of end points $\theta_0$ and $\theta_{\nparams+1}$ are determined by the above boundary conditions,
{\it i.e.} $\theta_0=0$ and $\theta_{\nparams+1}=1$.
A continuous control function $\Gamma(t)$ is obtained uniquely in terms of a cubic spline interpolation between the points $\theta_i$.

\subsubsection{Control target}

In principle, one could select a state fidelity as the control target.
Given the rapid growth of the Hilbert space with particle number $\nbosons$ and site number $\nsites$ however, the vast majority of quantum states would yield a negligibly small fidelity, and the state fidelity will fail to identify progress in the optimization.

With scalability in mind, we will therefore employ a different control target, defined in terms of the local particle numbers that can be experimentally observed for example with atomic gas microscopes~\cite{Bakr2009}.
Denoting the probability to observe exactly one atom at site $i$ by $p_i$, we will strive to minimize the average filling error
\be
\mathcal{E}=1-\frac{1}{\nsites}\sum_ip_i\ .
\ee
\new{Similarly to the infidelity, it yields the optimal value only for the target Mott-insulating state. 
In contrast to the state fidelity, however, states with an atom distribution similar to the Mott-insulating state, yield a close to optimal value.
The average filling error thus results in smoother control landscapes that allow algorithms to identify a reasonably broad peak around the target state, rather than an exponentially narrow peak as it would be the case with the state fidelity.
Moreover, the average filling error is less sensitive to readout errors. Those cause the fidelity to drop to the value of $0$, while they cause a reduction of the average filling error that decreases with the systems size}

\begin{figure}
  \includegraphics[width=0.49\textwidth]{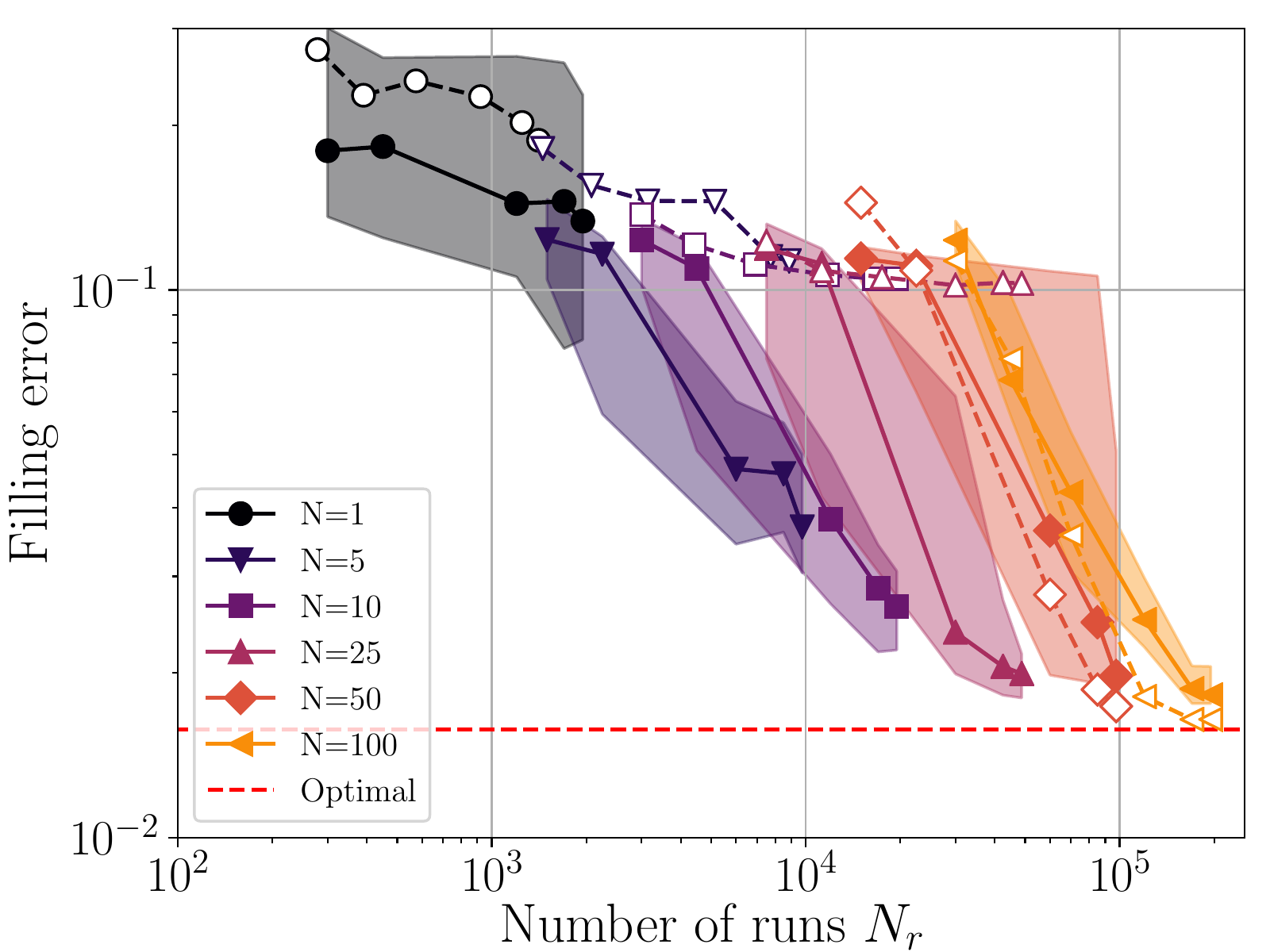}
	\caption{
   Convergence of the filling error $\mathcal{E}$ of a bosonic chain as function of the total number $\nq$ of runs.
Results obtained with binomial/Gaussian modeling are depicted with solid/empty symbols.
All optimizations are performed with a fixed number of repetitions $\nr$ of each measurement (specified in the legend) and a total number of iteration steps $M$ ranging from $250$ to $\num{2000}$. The dashed red line shows the result of the best solution found with $\mathcal{E}_{min} \approx 1.6\%$.}
	\label{fig:sfmi:bingauss}
\end{figure}

\subsubsection{Results}

Similarly to \new{the results presented in the previous example}, we will start with a comparison between Bayesian optimizations with binomial and Gaussian modeling.
Fig.~\ref{fig:sfmi:bingauss} depicts the result of this comparison in terms of the filling error $\mathcal{E}$ for different fixed number of repetitions per measurement, 
with all optimizations limited to a number $\ns=2000$ of iterative steps.

Reaching sub-percent filling errors is not possible for the fast protocol and the small number of control parameters ($\nparams=5$).
In order to gauge the performance of the optimizations with finite number of repetitions, we will compare them to optimizations without shot noise, {\it i.e.} perfectly resolved data.
In the latter case all the optimizations converge reliably to the same solution with a filling error $\mathcal{E}_{min} \approx 1.6\%$ that is shown in Fig.~\ref{fig:sfmi:bingauss} by a horizontal red dashed line.

Similarly to Sec.~\ref{sec:ghz}, binomial modeling (filled symbols) shows significantly better performance than its Gaussian counterpart (empty ones).
In particular with $25$ or fewer repetitions the Gaussian version failed to improve beyond the mediocre value of $\mathcal{E}=10\%$ of filling error, while the binomial ones converge steadily towards the supposed minimum $\mathcal{E}_{min}$.
Only with more frequent repetitions ($\nr=50$ and $\nr=100$ in Fig.~\ref{fig:sfmi:bingauss} ) does the Gaussian version manage to converge towards the low filling error $\mathcal{E}_{min}$.

\begin{figure}
    \includegraphics[width=0.49\textwidth]{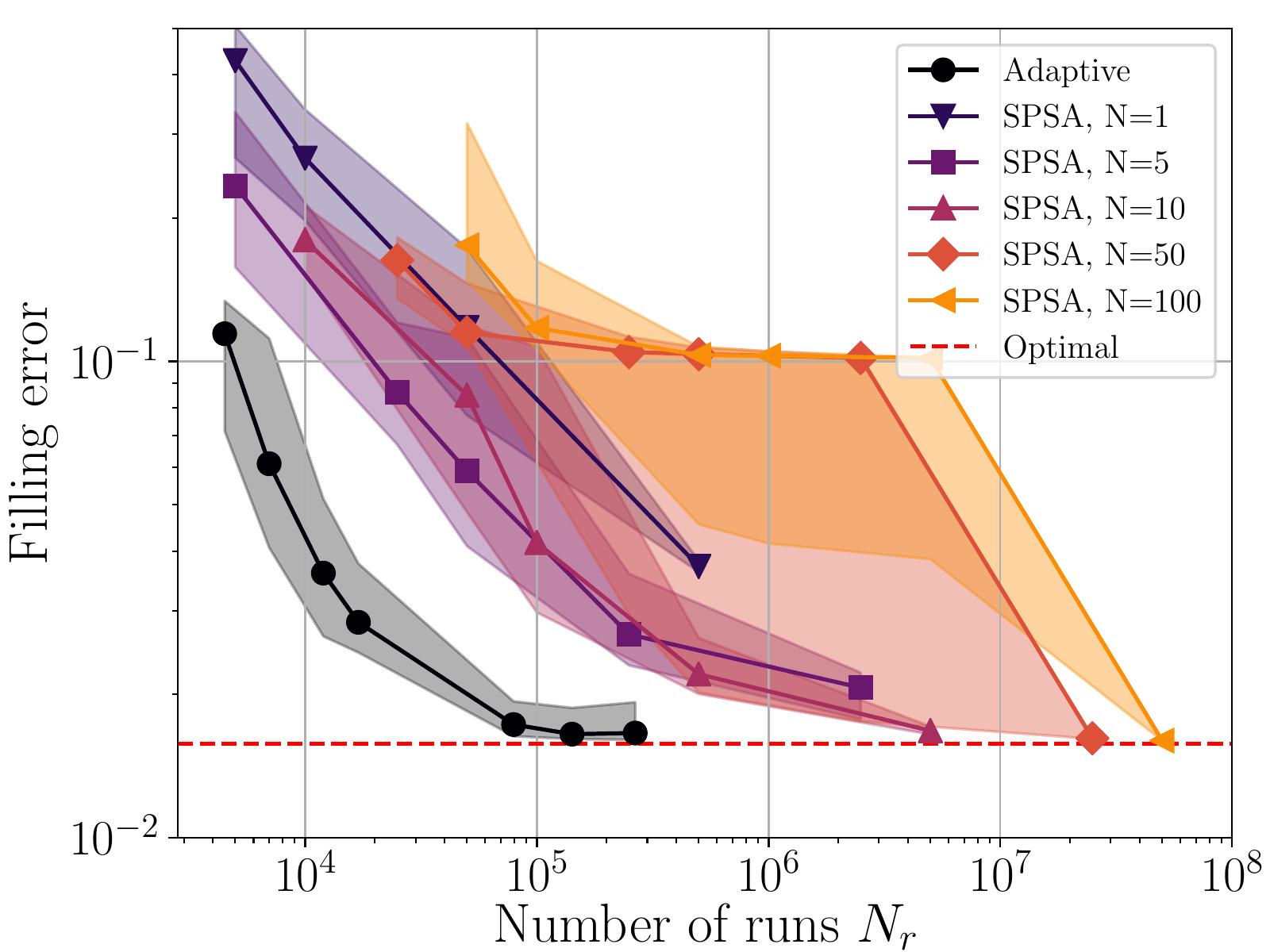}
    \caption{
    Optimization results for the adaptive strategy \new{(Sec.~\ref{sec:scalable})} 
    and SPSA with different number of repetitions $\nr$. The dashed red line indicates the lowest obtained filling error.}
    \label{fig:sfmi:resadaptive}
\end{figure}

Results of the adaptive strategy described in Sec.~\ref{sec:scalable} are presented in Fig.~\ref{fig:sfmi:resadaptive} (black curve). The first $4$ points of the curve pertain to the first stage with a small number of repetitions (here $\nr=10$).
Binomial modeling is used for up to to $\ns=1500$ steps.
After this, the parameter space is reduced (similarly to the example of GHZ states in Fig.~\ref{fig:adaptive}) and the number of repetitions is increased.
Results of this second round for $\num{1000}$ extra iterations steps and several choices for the increased number of repetitions $\nr=50,100,200$ are shown as the remaining three points of the curve. 
This strategy converges towards the supposed minimum $\mathcal{E}_{min}$ after a total number of $\nq\approx 10^5$ runs. 

Results obtained with the SPSA method for different numbers of fixed repetitions are depicted in color in Fig.~\ref{fig:sfmi:resadaptive}.
Consistently with the findings for GHZ states, depicted in Fig.~\ref{fig:adaptive}, the convergence with SPSA is substantially slower than convergence with Bayesian optimization and binomial modeling.
The optimizations with SPSA using $\nr=10$ and $\nr=100$ repetitions managed to converge close to $\mathcal{E}_{min}$, but did so after substantially more runs than with the adaptive method.
The optimizations with SPSA using $\nr=1$ and $\nr=5$ repetitions, even failed to converge to $\mathcal{E}_{min}$ within $2.5\times 10^5$ iterations.

\begin{figure}
    \includegraphics[width=0.49\textwidth]{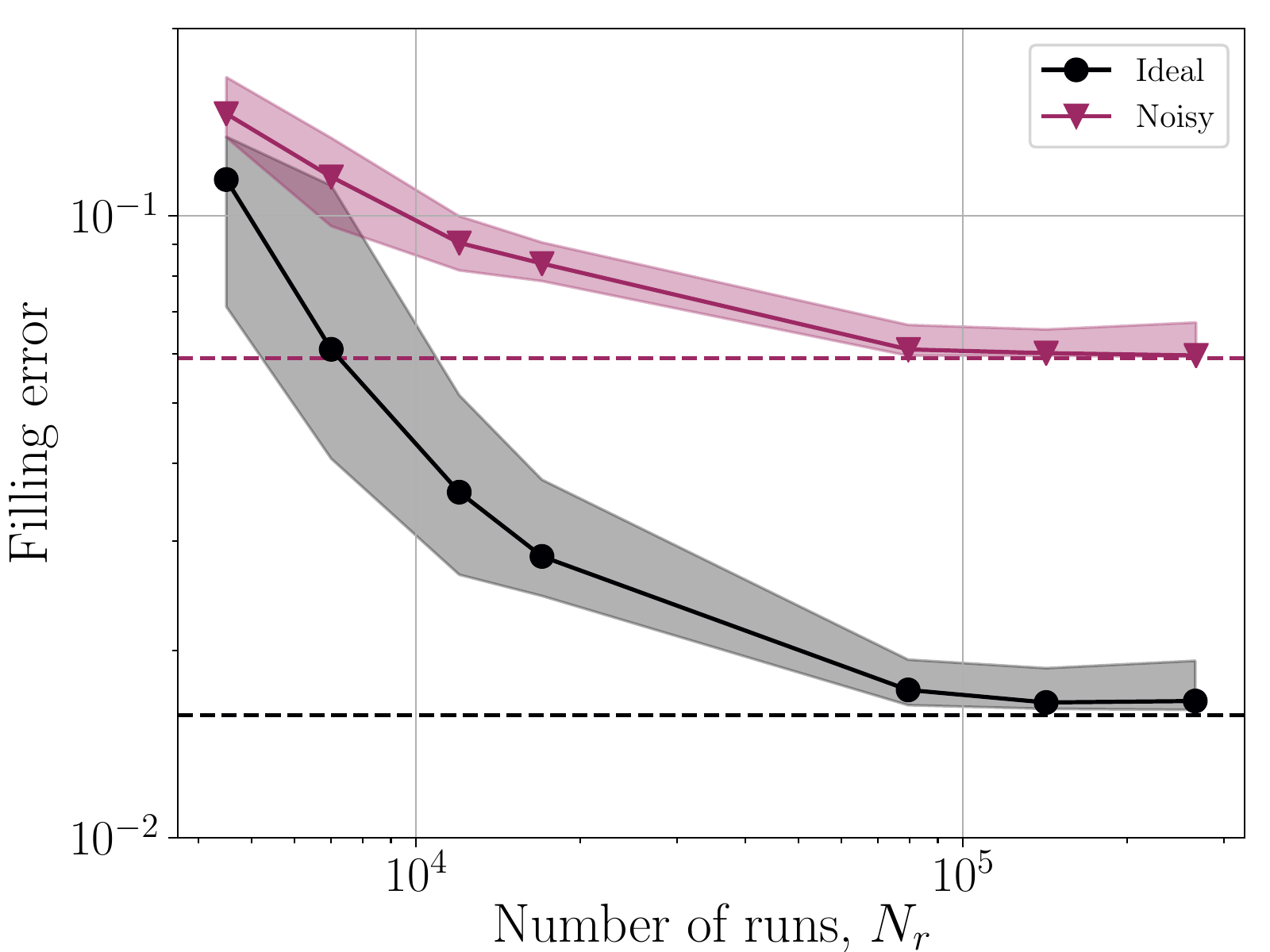}
    \caption{
 Convergence of optimizations in the presence and absence of fluctuations in the particle number $N_b$ (purple and black).
    In both cases the final filling errors are very close to the best filling errors obtained with simulations assuming perfectly accurate measurements (dashed lines).}
    \label{fig:noisy}
\end{figure}

Fig.~\ref{fig:noisy} gives evidence of the robustness of the framework against fluctuations in the number of trapped atoms.
It shows (in purple) convergence to low filling errors assuming a noise model with a $10\%$ chance for the system to be initialized with $4$ atoms or $6$ atoms respectively.
The filling error resulting from the best control found (obtained with perfect data without measurement noise) is depicted with a dashed line for reference. 
Corresponding data for the case without fluctuations in the particle number are included for comparison (in black).
As one can see, in both cases, there is very reliable convergence to an optimal solution.

Since we have found consistently that Bayesian optimization with binomial modeling outperforms all tested alternative approaches,
it seems in order to stress the aspects of this method that gives rise to this preferred behavior.
A general aspect of \bo is that at any step in the iterative procedure the full set of accumulated data is used in order to identify the next control parameters to probe, whereas gradient methods employ only data in the vicinity of one point of the control landscape.
As such, the scarce resource `data' is used in a more sustainable fashion and thus results in more accurate estimates of the control landscape. 
We explicitly verified that estimates of the control landscape based on measurement with few repetitions are substantially more accurate with Bayesian inference than with estimates in terms of relative frequencies of measurement outcomes.
This effect is particularly pronounced towards the maxima of the control landscape where probabilities of measurement results are close to the extreme values of `0' and `1'.

\new{
\subsection{Optimization on a NISQ device}\label{sec:ibmchips}

After verification of the optimization's convergence in theoretical simulations,
we will use the task of state-preparation on publicly available NISQ devices
namely the IBM Q Experience quantum chips~\cite{ibmq} as final demonstration.

The aim is to prepare the single-qubit target state
\begin{equation}\label{eq:ibmtgt}
	\ket{\Psi} = \cos \left(\frac{\pi}{8}\right) \ket{0}+e^{-i \pi / 4} \sin \left(\frac{\pi}{8}\right) \ket{1}\ , 
\end{equation}
using a parametrized circuit, composed of a rotation around the x-axis $R_x(\theta_1)=\exp[-i (\theta_1/2) \sigma_x]$ followed by a rotation around the z-axis $R_z(\theta_2)=\exp[-i (\theta_2/2) \sigma_z]$.

Similarly to Eq.~\eqref{eq:fid}, the fidelity of a state $\varrho$ with respect to the target state is recasted in terms of experimental observables as

\begin{equation} \label{eq:fidibm}
	F(\varrho) = \frac{1}{2}\left((1-\sqrt{2}) +\langle P_x\rangle - \langle P_y\rangle  + \sqrt{2}\langle P_z\rangle\right)\ ,
\end{equation}
where $P_i$ denotes the projector onto the eigenspace with positive eigenvalue
of the Pauli operator $\sigma_i$,
and $\langle P_i\rangle$ is a short-hand notation for $\mbox{Tr}(\varrho P_i)$.
\begin{table}
 \begin{center} \ra{1.3} \begin{tabular}{@{} m{4.9em} @{\hspace{0.7em}} l  @{\hspace{0.7em}} l c @{\hspace{0.7em}} l } \toprule 
 & \multicolumn{2}{c}{$\nq=300$} & & \multicolumn{1}{c}{$\nq=1500$} \\
\cline{2-3} \cline{5-5} 
 &\multicolumn{1}{c}{$\nr=1$} &\multicolumn{1}{c}{$\nr=5$} &&\multicolumn{1}{c}{$\nr=5$}\\ 
 \colrule 
bin. BO		& $0.966(0.011)$  	& $0.950(0.017)$  && $0.978(0.002)$ \\ 
Gauss. BO	& $0.852(0.087)$ 	& $0.943(0.020)$  && $0.958(0.034)$ \\ 
SPSA 		& $0.943(0.013)$	& $0.933(0.014)$  && $0.970(0.002)$ \\ 
 \botrule 
 \end{tabular} 
 \end{center} 
\caption{
\new{
State fidelities (Eq.\eqref{eq:fidibm}) resultant from optimizations with binomial \bo(bin. BO), Gaussian \bo(Gauss. BO) and SPSA on a NISQ device.
The fidelities are medians over several repetitions, and the values in brackets denote the uncertainty of these medians due to the finite number of repetitions.
}
}
\label{table:resqc}
\end{table}

Table~\ref{table:resqc} depicts the fidelities
of optimized gates,
based on binomial Bayesian optimization, Gaussian \bo and SPSA with $\nq=300$ and $\nq=1500$ runs.
While the optimizations are based on fidelities estimated with poor statistics, namely $\nr=1$ and $\nr=5$ repetitions,
the final fidelities reported are obtained with $\nr=\num{20000}$ repetitions.
The optimizations were performed during a time window of two weeks and were each repeated $5$ times,
except for the case with $N=5$ and $\nq=300$ (second column of Table.~\ref{table:resqc}) for which they were repeated $15$ times because of larger fluctuations.

Similarly to the discussion of the previous sections, the fidelities
in table~\ref{table:resqc} are medians over these repetitions.
In addition to variations due to the statistical nature of quantum mechanics and initializations of the search for optimal solutions,
there are also variations in the properties of the NISQ devices due to drift and re-calibration.
The resultant inaccuracy in the estimate of the state fidelities are depicted in brackets in table~\ref{table:resqc}.

Due to the limited accuracy of gates and readout in present NISQ devices, one can not expect to reach close-to-unit fidelities with any optimization method,
but a comparison of the fidelities resultant from the different control algorithms is meaningful.

As one can see in Table~\ref{table:resqc} 
the binomial version of \bo outperforms both SPSA and Gaussian \bo in all cases.
Gaussian \bo fails to reach high fidelities in the single-shot case ($\nr=1$),
but both binomial \bo and SPSA reach high fidelities.
Also the fluctuations in the fidelities obtained with the latter two control algorithms are lower than in the case of Gaussian Bayesian optimization.
Binomial \bo thus clearly outperforms Gaussian \bo as expected.
Furthermore, binomial \bo also outperforms SPSA. In the case of fewer runs of the circuits ($\nq=300$), this advantage is of the order of
several percent and,
only after optimizations with a larger number of repetitions  ($\nq=1500$) is the gap between the fidelities obtained with binomial \bo and SPSA smaller;
this highlights that optimizations with binomial \bo converge more quickly, making this approach more resource-efficient.}

\section{Conclusions}
While the description of measurement outcomes in terms of their full probability distribution
is extensively used in the context of \new{characterization of quantum systems~\cite{PhysRevA.84.052315,Shulman2014}},
quantum estimation and metrology~\cite{holevo2011probabilistic,Giovannetti2011,RevModPhys.89.035002},
it has hardly found applicability in the context of quantum \new{optimal} control.
The improvement resultant from accurate probabilistic modeling found here,
highlights that tools from statistical analysis can have similar impact also on quantum optimal control.

The ability to find close-to-optimal solutions with limited experimental data enabled by such statistical methodology, can advance technological development and precision experiments on a wide range of physical systems.
It offers a very resource-efficient pathway towards the optimal use of currently existing quantum hardware with $10$ to $100$ qubits;
the size of these systems makes theoretical modeling prohibitively expensive, and the
noisy character of the individual qubits calls for well-designed control sequences that prevent rapid accumulation of errors.
Given the availability of cheap resources for {\it classical} computation, it is essential to use them as much as possible to support the limited capabilities of near-term quantum hardware.
The present framework contributes directly to this goal in that it allows us to find optimal uses of quantum systems in terms of limited experimental data, but at the expense of increased computational overhead for \bo as compared to other control algorithms.
This trade-off, between the use of cheap classical computational resources versus expensive quantum mechanical ones can be tilted further in favor of one or the other.
Resorting to more accurate estimates of the landscape (e.g.~\cite{nickisch2008approximations} for a review) is likely to increase the efficiency with the quantum hardware, at the expense of classical computations.
On the other hand, the field of probabilistic machine learning~\cite{Ghahramani2015} provides fast approximate methods which could be incorporated into the framework.
For example probabilistic neural networks~\cite{NIPS2016_6117,NIPS2017_7219,DRLTRAP} and stochastic variational inference techniques~\cite{pmlr-v38-hensman15} hold the promise of increased efficiency in classical calculations, but this might imply slightly slower convergence towards the optimal control solution.

The applicability of the proposed methodology is by no means restricted to state preparation,
but includes for example the optimization of quantum gates in the presence of uncharacterized noise, or the direct realization of few qubit gates avoiding decomposition into more elementary gates in order to realize more complex quantum algorithms within limited coherence times.
Whereas optimization of gate fidelities will likely remain limited to systems comprised of few qubits,
there are also direct applications with large qubit registers.
One example is the variational quantum eigensolver~\cite{peruzzo2014variational,mcclean2016theory} aimed at finding the quantum state that minimizes an energy functional; the present method allows one to perform the optimization over a wide range of states without the need to accurately estimate their energy or the gradient of the landscape~\cite{2019arXiv190909083K,2019arXiv191001155S} during the search for the optimal state.
With the identification of ground states as a promising route towards the solutions of many classical problems such as the traveling salesman problem, the present methodology holds the potential to substantially advance the practical value of quantum systems for real life applications.

The improvement in performance shown here for the case of shot-noise in projective measurements is also not necessarily limited to this specific type of noise, but similar improvement should be expected for different types of noise such as environmental noise or noise in the actual control fields if their spectral properties or temporal correlations are taken into account appropriately.

With probabilistic modeling and \bo being active fields in mathematics and computer science, one should expect that the applicability and performance of similar methodologies will rapidly increase beyond what has been demonstrated in this work.\\

\section*{Acknowledgment}

We are indebted to stimulating discussions with Annie Pichery, Dominic Lennon, Robert L\"ow, Hongzheng Zhao, Jake Lishman, Ulrich Schneider and Zo\"{e} Holmes.
The project Theory-Blind Quantum Control {\it TheBlinQC} has received funding from the QuantERA ERA-NET Cofund in Quantum Technologies implemented within the European Union's Horizon 2020 Programme and from EPSRC under the grant EP/R044082/1. This work was supported through a studentship in the  Quantum  Systems  Engineering Skills and Training Hub at Imperial College London funded by EPSRC(EP/P510257/1).

\end{document}